\documentclass{aa}

\usepackage{graphicx}
\usepackage{txfonts}
\usepackage[dvipsnames]{xcolor}
\usepackage{tabularx}
\usepackage{chemformula}
\usepackage{physics}
\usepackage{scrextend}
\usepackage{makecell}
\usepackage[pdfpagelabels=false]{hyperref}	
\hypersetup{colorlinks=true,linkcolor=blue,citecolor=blue,filecolor=blue,urlcolor=blue,}
\usepackage{placeins}

\def\figurenum#1{\def\thefigure{#1}\let\currentlabel\thefigure}

\begin{document} 

\title{Information content of JWST spectra of WASP-39b}
\titlerunning{Information content of JWST spectra of WASP-39b}
\authorrunning{Lueber \& Novais et al.}

\author{Anna Lueber\inst{\thanks{Both authors contributed equally to the project},1}
         \and Aline Novais\inst{\star,2,1}
         \and  Chloe Fisher\inst{3}
         \and  Kevin Heng\inst{1,4,5,6}}
\institute{Ludwig Maximilian University, Faculty of Physics, University Observatory, Scheinerstr. 1, Munich D-81679, Germany\\
\email{anna.lueber@physik.lmu.de}
\and Valongo Observatory, Federal University of Rio de Janeiro, Ladeira do Pedro Antonio, 43, 20080-090, Rio de Janeiro, Brazil\\
\email{aline12@ov.ufrj.br}
\and Astrophysics, University of Oxford, Denys Wilkinson Building, Keble Road, Oxford, OX1 3RH, United Kingdom
\and ARTORG Center for Biomedical Engineering Research, University of Bern, Murtenstrasse 50, CH-3008, Bern, Switzerland
\and University College London, Department of Physics \& Astronomy, Gower St, London, WC1E 6BT, United Kingdom
\and University of Warwick, Department of Physics, Astronomy \& Astrophysics Group, Coventry CV4 7AL, United Kingdom
}

\date{Received ; accepted}

\abstract {The era of \textit{James Webb Space Telescope} (JWST) transmission spectroscopy of exoplanetary atmospheres commenced with the study of the Saturn-mass gas giant WASP-39b as part of the Early Release Science (ERS) program.  WASP-39b was observed using several different JWST instrument modes (NIRCam, NIRISS, NIRSpec G395H and NIRSpec PRISM) and the spectra were published in a series of papers by the ERS team.}
{The current study examines the information content of these spectra measured using the different instrument modes, focusing on the complexity of the temperature-pressure profiles and number of chemical species warranted by the data.  We examine if the molecules \ch{H2O}, \ch{CO}, \ch{CO2}, \ch{K}, \ch{H2S}, \ch{CH4}, and \ch{SO2} are detected in each of the instrument modes.}
{Two Bayesian inference methods are used to perform atmospheric retrievals: the standard nested sampling method, as well as the supervised machine learning method of the random forest (trained on a model grid).  For nested sampling, Bayesian model comparison is used as a guide to identify the set of models with the required complexity to explain the data.} 
{Generally, non-isothermal transit chords are needed to fit the transmission spectra of WASP-39b, although the complexity of the temperature-pressure profile required is mode-dependent.  The minimal set of chemical species needed to fit a spectrum is mode-dependent as well, and also depends on whether grey or non-grey clouds are assumed. When a non-grey cloud model is used to fit the NIRSpec G395H spectrum, it generates a spectral continuum that compensates for the water opacity. The same compensation is absent when fitting the non-grey cloud model to the NIRSpec PRISM spectrum (which has broader wavelength coverage), suggesting that it is spurious. The interplay between the cloud spectral continuum and the water opacity determines if sulphur dioxide is needed to fit either spectrum.}
{The inferred elemental abundances of carbon and oxygen and the carbon-to-oxygen (C/O) ratios are all mode- and model-dependent, and should be interpreted with caution. Bayesian model comparison does not always offer a clear path forward for favouring specific retrieval models (e.g. grey versus non-grey clouds) and thus for enabling unambiguous interpretations of exoplanet spectra.}
\keywords{Planets and satellites: atmospheres - Planets and satellites: composition - Techniques: spectroscopic - planets and satellites: individual: WASP-39b}

\maketitle

\section{Introduction}
\label{sect:intro}

The hot Jupiter WASP-39b’s optimal combination of low surface gravity ($\log{g}\approx 2.6$ in cgs units; \citealt{Faedi2011A&A...531A..40F}) and somewhat high equilibrium temperature (about 1100 K; \citealt{Faedi2011A&A...531A..40F}), which collectively yields a large atmospheric pressure scale height, makes it an optimal target for transmission spectroscopy.  For these reasons, it was chosen as one of the targets of the JWST Transiting Exoplanet ERS program \citep{Stevenson2016PASP..128i4401S, Bean2018PASP..130k4402B, Ahrer2023Natur.614..649J}.  The ERS team has measured transmission spectra of WASP-39b using the NIRCam \citep{Ahrer2023Natur.614..653A}, NIRISS \citep{Feinstein2023Natur.614..670F}, NIRSpec G395H \citep{Alderson2023Natur.614..664A}, and NIRSpec PRISM \citep{Ahrer2023Natur.614..649J, Rustamkulov2023Natur.614..659R} instrument modes of JWST.

WASP-39b was previously observed using the Wide Field Camera 3 (WFC3) onboard the Hubble Space Telescope, where the detection of water was reported \citep{Wakeford+18}.  Due to the limited wavelength range covered by WFC3 (0.8 to 1.7~$\mu$m), it was not possible to make definitive statements on other molecular species.  Sodium was also detected using the Space Telescope Imaging Spectrograph (STIS) onboard Hubble, which covered the wavelength range of 0.29 to 1.025 $\mu$m \citep{Fischer+16}. The JWST spectra of WASP-39b collectively cover a wavelength range of 0.5 to 5.5 microns, which contains the spectral features of all of the simple carbon, hydrogen, and oxygen carriers (\ch{H2O}, \ch{CO}, \ch{CO2}, \ch{CH4}), as well as that of potassium.  The tentative detection of sulphur dioxide (\ch{SO2}) was also reported \citep{Rustamkulov2023Natur.614..659R}, which is believed to be a photochemical byproduct of hydrogen sulphide (\ch{H2S}) and evidence for the presence of photochemistry \citep{Tsai2023Natur.617..483T}.

Collectively, the ERS team has reported the detection of carbon dioxide \citep{Ahrer2023Natur.614..649J, Alderson2023Natur.614..664A, Rustamkulov2023Natur.614..659R}, water \citep{Alderson2023Natur.614..664A, Ahrer2023Natur.614..653A, Feinstein2023Natur.614..670F, Rustamkulov2023Natur.614..659R}, carbon monoxide \citep{Rustamkulov2023Natur.614..659R}, sulphur dioxide \citep{Alderson2023Natur.614..664A, Rustamkulov2023Natur.614..659R}, sodium \citep{Rustamkulov2023Natur.614..659R}, and potassium \citep{Feinstein2023Natur.614..670F}, as well as a non-detection of methane \citep{Ahrer2023Natur.614..649J, Ahrer2023Natur.614..653A, Rustamkulov2023Natur.614..659R}.  Carbon dioxide was also detected by applying a cross correlation method to the NIRSpec G395H spectrum \citep{Esparza-Borges2023ApJ...955L..19E}.  Fitting forward models to the spectra of the different instrument modes yields super-solar metallicities to varying degrees \citep{Alderson2023Natur.614..664A, Ahrer2023Natur.614..653A, Feinstein2023Natur.614..670F, Rustamkulov2023Natur.614..659R}.  However, whether the C/O ratio is sub-solar \citep{Alderson2023Natur.614..664A, Ahrer2023Natur.614..653A, Feinstein2023Natur.614..670F}, solar \citep{Alderson2023Natur.614..664A} or super-solar \citep{Rustamkulov2023Natur.614..659R} depends on the instrument mode being considered.  The mode and model dependence of the inferred atmospheric metallicity and C/O ratio is confirmed by the present study.

The richness of information encoded in these JWST spectra of WASP-39b, as well as the differences in the inferred metallicities and C/O ratios motivate a thorough investigation using a Bayesian framework of inference.  In the field of exoplanetary atmospheres, this is known as atmospheric retrieval \citep{MadhusudhanSeager2009ApJ...707...24M}, which solves the inverse problem of inferring chemical abundances and atmospheric properties from fitting a measured spectrum (for a review, see \citealt{BarstowHeng2020SSRv..216...82B}).  In the current paper, we study the JWST spectra of WASP-39b using a suite of atmospheric retrievals.  We implement these retrievals in two ways: first, using the standard approach of nested sampling \citep{Skilling2006AIPC..872..321S}; second, using the supervised machine learning method of the random forest \citep{Marquez-Neila2018NatAs...2..719M}.  Questions we wish to address include:

\begin{enumerate}
    \item Are isothermal transit chords sufficient to fit JWST transmission spectra?  If not, what is the minimum level of complexity (non-isothermal behaviour) needed to produce a best fit?
    \item Which molecules may we declare to be detected?
    \item Are grey or non-grey clouds required to fit the spectrum?  And how do they interact with other components of the model?
    \item Are any of our inferences dependent on the instrument mode being considered?
\end{enumerate}

We anticipate that a family of models, rather than a single model, will be able to fit each spectrum. To implement Occam’s Razor and penalise models that are too complex (given the data quality), we implement Bayesian model comparison \citep{Trotta2008ConPh..49...71T}, which is a natural outcome of the nested sampling algorithm \citep{Skilling2006AIPC..872..321S}.  As we report, there are instances where Bayesian model comparison does not allow us to clearly select a single retrieval model for interpreting the data.

In Sect.~\ref{sect:methods}, we describe our methodology including the data curated and the retrieval methods implemented.  In Sect.~\ref{sect: results}, we present the outcomes of our suites of retrievals.  In Sect.~\ref{sect: discussion}, we summarise our findings and discuss their implications.

\begin{figure*}[ht!]
    \centering
    \includegraphics[width=\textwidth]{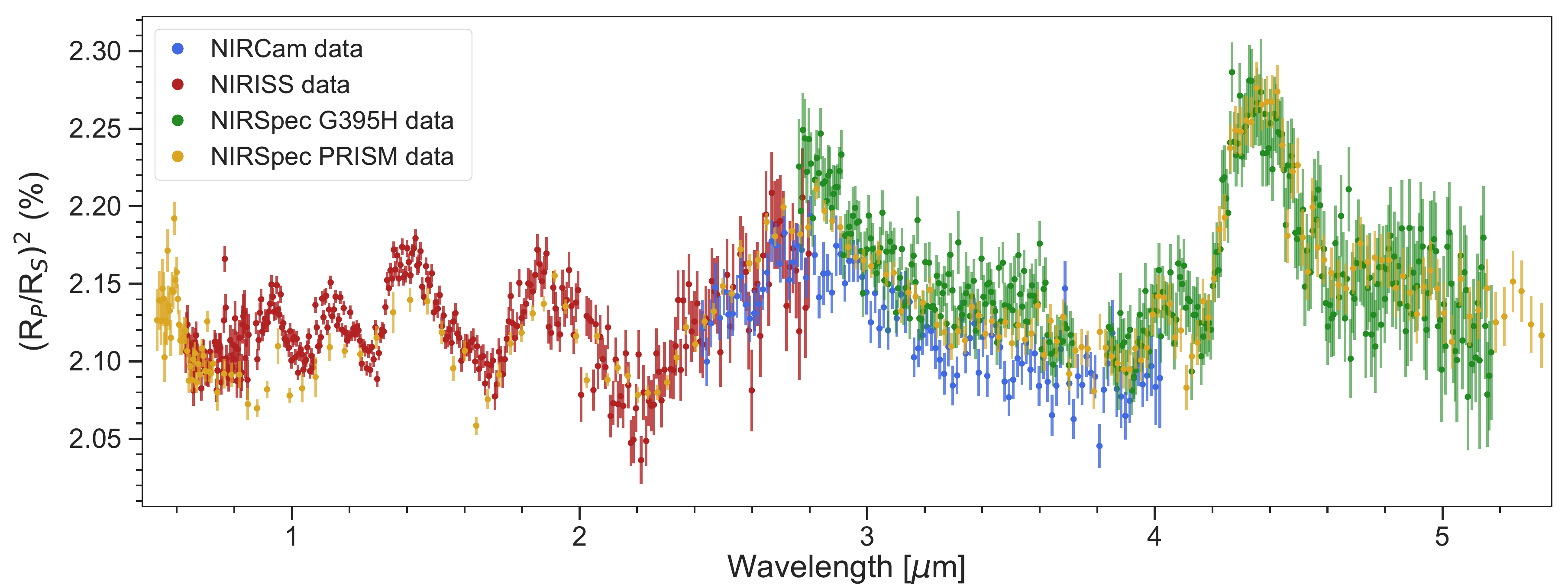}
    \caption{Curated JWST ERS transmission data of WASP-39b from the following four instruments: NIRCam (blue), NIRISS (red), NIRSpec G395H (green), and NIRSpec PRISM (yellow). The data were originally published by \citet{Ahrer2023Natur.614..653A}, \citet{Feinstein2023Natur.614..670F}, \citet{Rustamkulov2023Natur.614..659R}, and \citet{Alderson2023Natur.614..664A}, respectively.}
    \label{fig:Spectra}
\end{figure*}

\section{Methodology}
\label{sect:methods}

\subsection{Spectral sample}
\label{sect:sample}

WASP-39b is a 0.28 M$_\mathrm{J}$ and 1.27 R$_\mathrm{J}$ Saturn-like exoplanet with an equilibrium temperature of about 1100 K, orbiting a G7 V star with a mass of 0.93 M$_\odot$ and radius of 0.895 R$_\odot$ \citep{Faedi2011A&A...531A..40F}. No new data are presented in the current study.  We analyse the four spectra previously published by \cite{Ahrer2023Natur.614..653A} (NIRCam, 2.4--4.0 $\mu$m), \cite{Feinstein2023Natur.614..670F} (NIRISS, 0.6--2.8 $\mu$m), \cite{Alderson2023Natur.614..664A} (NIRSpec G395H, 2.7--5.2 $\mu$m), and \cite{Rustamkulov2023Natur.614..659R} (NIRSpec PRISM, 0.5--5.5 $\mu$m).  We make no attempt at re-reducing or combining these spectra, or correcting for detector offsets between them.  The four spectra used in the current study are displayed in Fig.~\ref{fig:Spectra}.

\subsection{Atmospheric retrieval techniques}

\subsubsection{Nested sampling}
\label{sect:atmosphericretrieval}

\begin{figure*}[ht]
    \centering
    \includegraphics[width=\textwidth]{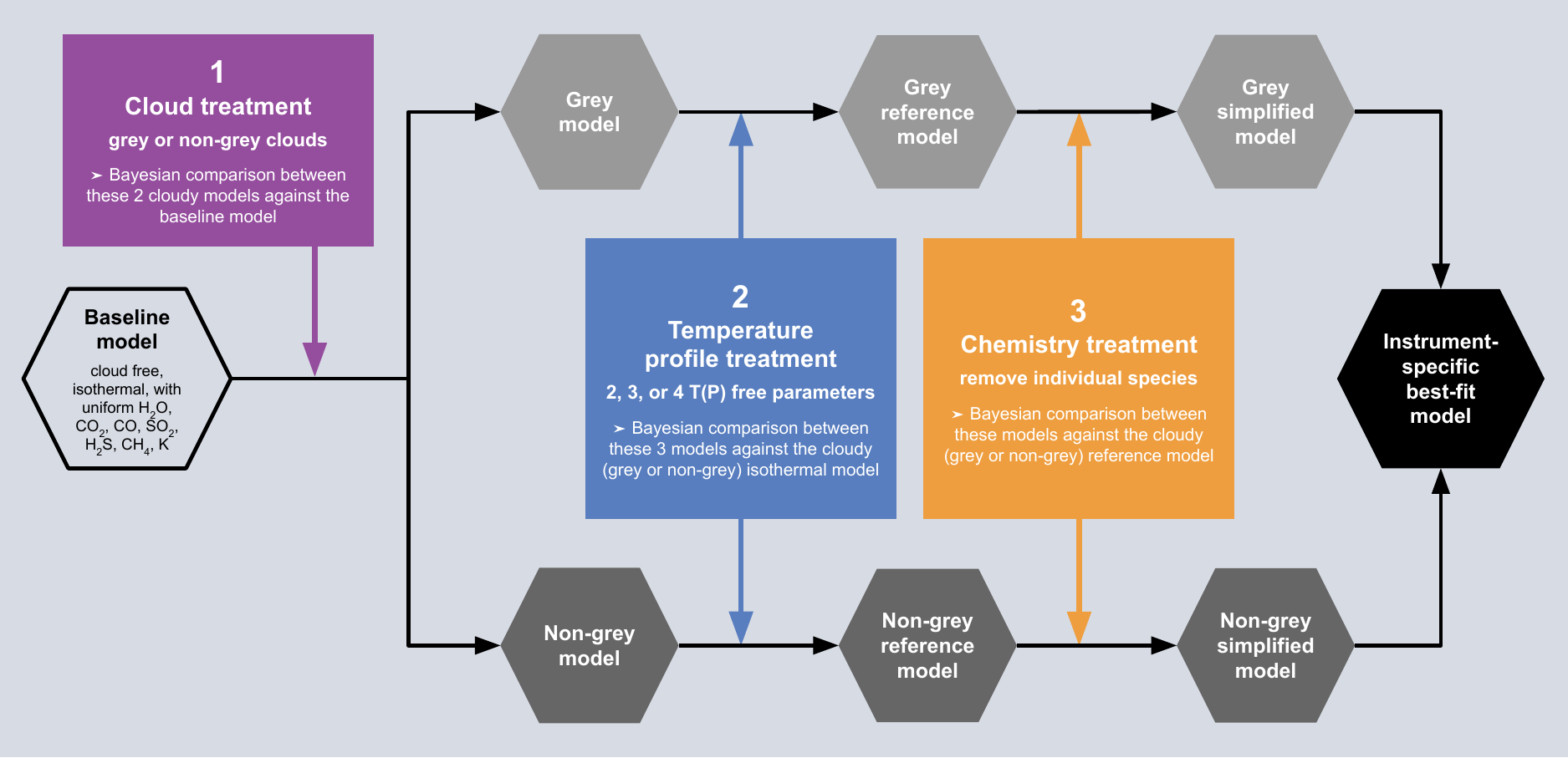}
    \caption{Schematic describing our suite of nested sampling retrievals, where models with grey and non-grey clouds are tested separately for variations in the temperature-pressure profile, and number of atoms and molecules considered.}
    \label{fig:Heliosr2_Scheme}
\end{figure*}

For standard Bayesian retrievals, we use the open-source, GPU-accelerated \texttt{Helios-r2} code \citep{Kitzmann2020ApJ...890..174K}\footnote{\href{https://github.com/exoclime}{https://github.com/exoclime} \label{ESP}}.  It is based on the nested sampling technique \citep{Skilling2006AIPC..872..321S} and implements the \texttt{MULTINEST} algorithm \citep{Feroz2008MNRAS.384..449F, Feroz2009MNRAS.398.1601F}.  \texttt{Helios-r2} is capable of fitting both transmission and emission spectra, and has the option to include grey and non-grey clouds (see \citealt{Kitzmann2020ApJ...890..174K} and \citealt{Lueber2022ApJ...930..136L} for details).  The one-dimensional, plane-parallel model atmosphere consists of 99 layers (100 levels) spanning 10 bar to 1 $\mu$bar. The temperature-pressure profile is described by a finite element approach that ensures smoothness and continuity.  In the current study, it is parameterised by between one (isothermal) and four parameters, where the required complexity is data-driven and guided by Bayesian model comparison \citep{Trotta2008ConPh..49...71T}. 

The marginalised likelihood, which is also termed the Bayesian evidence, is computed for each retrieval model.  The ratio of Bayesian evidences between a pair of retrieval models is termed the ``Bayes factor'' ${\cal B}$.  It may, in principle, allow one to identify the model that best explains the data with the appropriate level of complexity \citep{Trotta2008ConPh..49...71T}.  For the convenience of the reader, we summarise some key points from Table 2 of \cite{Trotta2008ConPh..49...71T}: when $\ln{\cal B}<1$, both models explain the data equally well.  When $\ln{\cal B} \ge 5$, the model with the higher Bayesian evidence is strongly favoured.  When $\ln{\cal B} \ge 2.5$, it is only moderately favoured.

The open-source \texttt{HELIOS-K} opacity calculator \citep{Grimm2015ApJ...808..182G, Grimm+21} was used to convert spectroscopic line lists into absorption cross sections or opacities (cross sections per unit mass) of atoms and molecules. We utilise the line lists for \ch{H2O} \citep{Polyansky+18}, \ch{CO2} \citep{Tashkun+11}, \ch{CO} \citep{Li+15}, \ch{SO2} \citep{Underwood2016MNRAS.459.3890U}, \ch{H2S} \citep{Azzam2016MNRAS.460.4063A}, and \ch{CH4} \citep{YurchenkoTennyson14}.  The resonant line wing shapes of the alkali metals (\ch{K} and \ch{Na}) are taken from \cite{Allard2016A&A...589A..21A, Allard2019A&A...628A.120A}.  Collision-induced absorption (CIA) by \ch{H2}–\ch{H2} \citep{Abel2011JPCA..115.6805A} and \ch{H2}–\ch{He} \citep{Abel2012JChPh.136d4319A} pairs are included.  For a review of spectroscopic databases, please see \cite{Tennyson2017MolAs...8....1T}.  The opacities are publicly available on the DACE opacity database \citep{Grimm+21}\footnote{\href{https://dace.unige.ch}{https://dace.unige.ch} \label{DACE}}.

The grey cloud model simply assumes a single, constant opacity.  Our non-grey cloud model is taken from equation (32) of \cite{KitzmannHeng18} and is calibrated on refractive indices measured in the laboratory of a library of different aerosol compositions.  Based on an empirical description of Mie theory, it smoothly connects extinction by small and large particles, which correspond to Rayleigh scattering and grey extinction, respectively \citep{Pierrehumbert2010ppc..book.....P}. Spherical particles of a single size are assumed.  The parameters of our non-grey cloud model include the cloud-top pressure $P_{\rm cloudtop}$, the optical depth referenced to 1 $\mu$m, a composition parameter $Q_0$ (which determines the value of $2\pi r_{\rm cloud}/\lambda$ where the transition from small- to large-particle extinction occurs), the particle radius $r_{\rm cloud}$, and the index of the slope corresponding to small-particle scattering $a_0$ (which is exactly 4 for Rayleigh scattering). For both the grey and non-grey cloud models, the cloud is assumed to be semi-infinite, meaning the cloud-bottom pressure extends to the bottom of the model domain.\footnote{In practice, this is implemented in \texttt{HELIOS-r2} by setting the cloud-bottom pressure to be $10^7$ times that of the cloud-top pressure.}

The parameter space to be explored is vast, since we have to consider multiple atomic and molecular species (in various combinations), non-isothermal temperature-pressure profiles and grey versus non-grey clouds.  Fig.~\ref{fig:Heliosr2_Scheme} describes how we conduct our parameter explorations by examining the treatment of the temperature-pressure profile and chemistry (number of species included).

\begin{figure}[ht!]
\centering
\includegraphics[width=0.9\columnwidth]{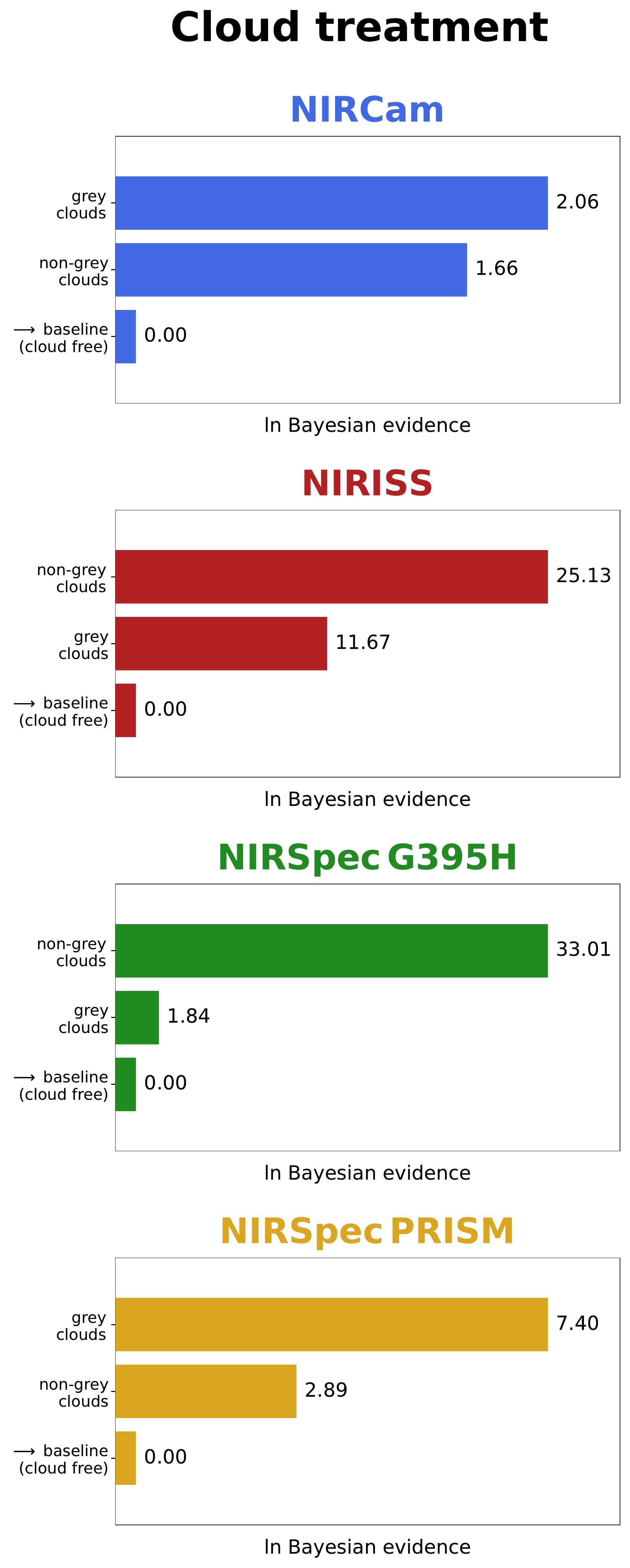}
\caption{Bayesian model comparison between cloudfree and cloudy retrievals, corresponding to the first step of the schematic shown in Fig.~\ref{fig:Heliosr2_Scheme}. The Bayesian evidence relative to the cloudfree model is computed and its natural logarithm is displayed. For all four JWST instruments, there are significant increases in the Bayesian evidence when cloudy models are used to fit the spectra.}
\label{fig:bayes_clouds}
\end{figure}

Our baseline model is cloudfree, isothermal, and contains vertically uniform abundances for \ch{H2O}, \ch{CO2}, \ch{CO}, \ch{SO2}, \ch{H2S}, \ch{CH4}, and \ch{K}. For simplicity, the \ch{Na} abundance is not treated as a fitting parameter and is instead assumed to be given by the solar elemental abundance ratio between it and \ch{K}. However, NIRSpec PRISM retrievals present an exception where the line centre is accessible, allowing \ch{Na} to be treated as an independent fitting parameter. The cloudfree model is then compared to models with grey versus non-grey clouds.  For all four spectra, cloudfree models are disfavoured, to different degrees, via Bayesian model comparison \citep{Trotta2008ConPh..49...71T}, yielding values for the logarithm of the Bayes factor from 1.66 to 33.01, as documented in Fig. \ref{fig:bayes_clouds}.  Since one of our motivations is to understand how the cloud model interacts with other components of the retrieval, we run two parallel sets of retrievals for grey and non-grey clouds (Fig.~\ref{fig:Heliosr2_Scheme}).

For each set of retrievals (with either grey or non-grey clouds), we next explore the effects of varying the complexity of the temperature-profile.  We use Bayesian model comparison to guide our choice of temperature-pressure profile, which has between 1 and 4 fitting parameters.  If a subset of these models (with different temperature-pressure profiles) are associated with $\ln{\cal B}<1$, then we apply Occam's Razor and select the model with the least number of parameters describing the temperature-pressure profile. We term this the ``reference model''.

After the temperature-pressure profile has been chosen, we remove the chemical species one at a time and record the Bayesian evidence.  If the Bayesian evidence is higher when a species is removed, then we exclude it from the model regardless of whether the logarithm of the Bayes factor is greater or less than unity.  If the Bayesian evidence is lower when a species is removed \textit{and} the logarithm of the Bayes factor exceeds unity, then we include this species in the model. By considering each of the seven species in turn (eight species for the PRISM retrieval), we are able to identify the model with the least number of chemical species required to fit the data.  We term this model the ``simplified model''.


The Bayesian framework used within \texttt{Helios-r2} requires us to define prior distributions for all fitting parameters of the forward model. All parameters and their priors may be found in Table~\ref{tab:helios_r2_prior_ranges}.  In particular, the assumed prior distributions of the stellar radius $R_s$, planetary white-light radius $R_p$, and planetary surface gravity $g$ are based on the measurements of \citet{Mancini+18}.

\begin{table}[ht]
\centering
\caption{Summary of retrieval parameters and the assumptions for their prior distributions.}
\label{tab:helios_r2_prior_ranges}
\resizebox{0.5\textwidth}{!}{
\begin{tabular}{lcccc}
\hline
Parameter & Symbol & Prior range & Distribution & Units \\ \hline
Temperature & $T$ & [500, 3000] & Uniform & K \\
\makecell[cl]{Slope between two adjacent \\ temperature nodes$^{*}$} & $b_{i=1\dots4}$    & [0.1, 3.0] & Uniform & - \\
Molecular abundances & $X_{i}$ & [10$^{-12}$, 10$^{-1}$] & Log-uniform & - \\
Planetary radius & $R_p$ & 1.279$\pm$0.051 & Gaussian & R$_{\rm J}$ \\
Stellar radius & $R_S$ & 0.939$\pm$0.030 & Gaussian & R$_\odot$ \\
Planetary surface gravity & ${\rm log} \, g$ & 2.629$\pm$0.051 & Gaussian & cm s$^{-2}$ \\ 
\hline
\textit{Grey clouds} &  & & &   \\
Cloud-top pressure & P$_{\rm cloudtop}$ & [10$^{-6}$, 10] & Log-uniform & bar \\
Optical depth & $\tau_{\rm cloud}$ & [10$^{-5}$, 10$^{3}$] & Log-uniform & - \\
\hline
\textit{Non-grey clouds} &  & & &   \\
Cloud-top pressure & P$_{\rm cloudtop}$ & [10$^{-6}$, 10] & Log-uniform & bar \\
Reference optical depth & $\tau_{\rm cloud}$ & [10$^{-5}$, 10$^{3}$] & Log-uniform & - \\
Composition parameter & Q$_0$ & [1, 100] & Uniform & - \\
Index & a$_0$ & [3, 6] & Uniform & - \\
Spherical cloud particle radius & $r_{\rm cloud}$ & [10$^{-7}$, 10$^{-1}$] & Log-uniform & cm \\
\hline
\end{tabular} }
{\raggedright $^{*}$ \footnotesize Only for the non-isothermal profile, where $b_i$ is interpreted as the slope between two adjacent temperature nodes.  More details may be found in \cite{Kitzmann2020ApJ...890..174K}. \par}
\end{table}

\subsubsection{Random forest}
\label{sect: MLretrieval}

We also perform atmospheric retrieval using the supervised machine learning method of the random forest \citep{Ho1998random, Breiman2001random} implemented via the code \texttt{HELA} \citep{Marquez-Neila2018NatAs...2..719M}\footref{ESP}.  It uses the pre-computed grid of model spectra published by \cite{Crossfield2023ApJ...952L..18C} as a training set for performing inference in the framework of Approximate Bayesian Computation \citep{Sisson2018handbook}. This model grid was computed using the \texttt{VULCAN} photochemical kinetics code \citep{Tsai+17, Tsai+21} and the \texttt{petitRadTrans} radiative transfer code for generating transmission spectra \citep{Molliere+19}. It considers metallicities from solar to $100\times$ solar (by varying the elemental abundances of \ch{C}, \ch{O}, and \ch{S}), which amounts to 1331 transmission spectra in total. It implements a spectral resolution of $\sim$1000 between 1 to 25~$\mu$m. This wavelength range implies that part of the NIRISS and NIRSpec PRISM spectra are not analysed.

To use the model grid as a training set, we bin the model spectrum (in terms of transit depth $D$) according to the wavelength bins of each measured spectrum.  The specific details of the binning differ between the spectra from the four instrument modes.  Within each bin, there is a measured uncertainty $\delta$.  We use $\delta/D$ as the standard deviation of a Gaussian from which we randomly draw uncertainties for the model spectrum.  This uncertainty is then added or subtracted from the binned model transit depth \citep{Oreshenko2020AJ....159....6O, Lueber2023ApJ...954...22L}.  In this manner, even the same model spectrum may generate multiple realisations of model spectra that include noise.  Using this approach, we generate a training set (with noise) of 2662 model spectra using the \cite{Crossfield2023ApJ...952L..18C} grid. A total of 3000 regression trees with no restriction on the maximum tree depth were used. Instead, each tree grows by splitting the space of the models until the decrease in variance in the associated parameters of further splits is smaller than a fractional value of 0.01.

\section{Results}
\label{sect: results}
\subsection{Nested sampling retrievals}


\subsubsection{Complexity of temperature-pressure profile}
\label{sect: TP_treatment}

\begin{figure*}
\centering
\includegraphics[width=\textwidth]{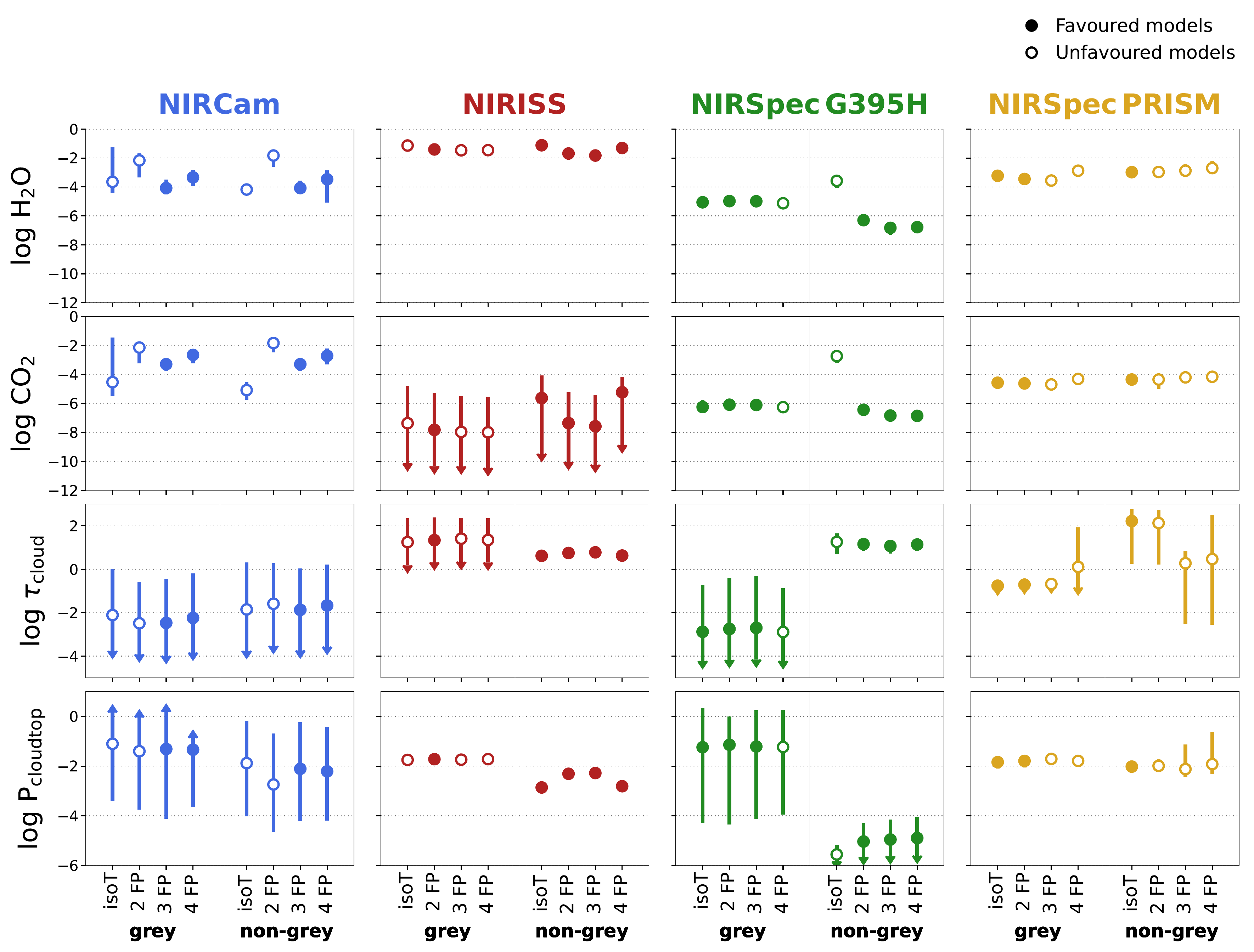}
\vspace{-0.4cm}
\caption{Comparing the median values of posterior distributions of selected parameters from retrievals performed on the four different spectra.  Arrows indicate upper or lower limits.  For each spectrum, a suite of four retrievals with different complexity of the temperature-pressure profile is performed.  Isothermal profiles are denoted by ``isoT'', while ``FP'' means that the temperature-pressure profile is parameterised by F fitting/free parameters.  Separate suites of retrievals for grey versus non-grey cloud models are performed.  Within each suite, filled circles are models where the logarithm of the Bayes factor is less than unity, implying that all of these models explain the data equally well.  Within this subset of models indicated by filled circles, we choose the temperature-pressure profile that uses the least number of parameters in our subsequent explorations.} 
\label{fig:Quest}
\end{figure*}

Fig.~\ref{fig:Quest} shows the outcomes of a suite of retrievals that explore the complexity of the temperature-pressure profile used, while including all seven chemical species (eight for PRISM retrievals) previously described.  To explore the sensitivity of outcomes to the choice of cloud model, separate sub-suites of retrievals are performed using grey versus non-grey cloud models.  Let the Bayes factor be ${\cal B}$ and its natural logarithm be $\ln{\cal B}$.  In Fig.~\ref{fig:Quest}, the filled circles indicate the subset of models where $\ln{\cal B}<1$ (relative to the model with the highest Bayesian evidence).  We term this subset the ``favoured models''.  

\begin{figure*}[ht]
    \centering
    \includegraphics[width=0.95\linewidth]{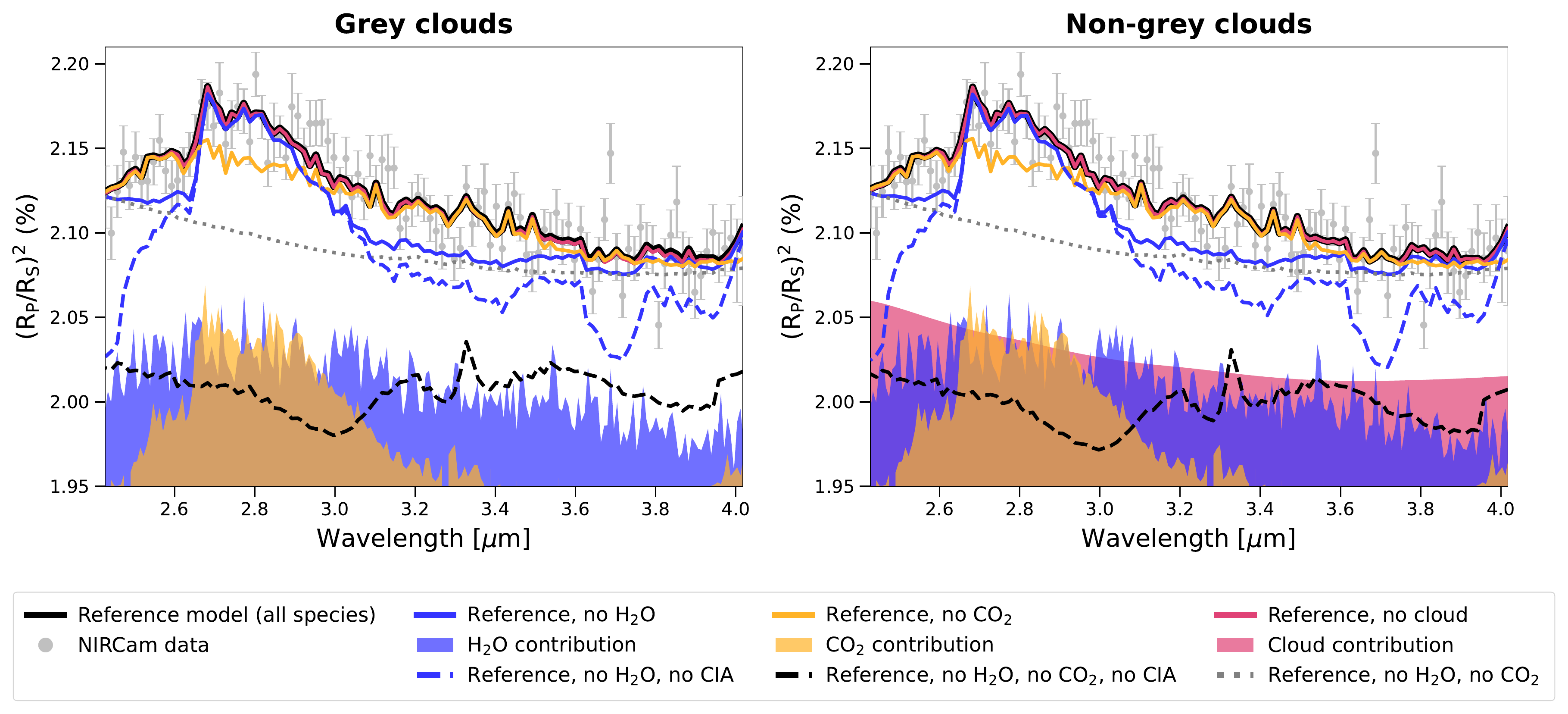}
    \caption{Fitting the model containing all chemical species (termed the ``reference model'') to the NIRCam spectrum (grey data points) previously published by \cite{Ahrer2023Natur.614..653A}. Corresponding to our findings in Fig.~\ref{fig:Quest}, the 3-parameter temperature-pressure profile is used and grey (left panel) versus non-grey (right panel) are considered in separate retrievals.  In each panel, post-processing is performed to isolate the contribution of water, carbon dioxide, clouds, and collision-induced absorption (CIA) due to hydrogen and helium.  The natural logarithm of the Bayes factor between the retrievals with grey and non-grey clouds is 0.7, implying that both grey and non-grey cloud models explain the data equally well because both clouds are optically thin.}
    \label{fig:postprocessed_NIRCam}
\end{figure*}

The water and carbon dioxide abundances retrieved from NIRCam spectra depend on the complexity of the temperature-pressure profile adopted and not on the choice of grey versus non-grey model.  When isothermal transit chords are assumed, cloudy models are weakly preferred for fitting NIRCam spectra with $\ln{\cal B}=2.06$ and $\ln{\cal B}=1.66$ for grey and non-grey clouds, respectively (Fig. \ref{fig:bayes_clouds}). For the isothermal retrieval, the optical depth for the grey cloud is $\log{\tau_{\rm cloud}}=-2.11_{-1.65}^{+2.13}$.  When non-isothermal temperature-pressure profiles are used the clouds remain optically thin (Fig.~\ref{fig:Quest}).  Fig.~\ref{fig:postprocessed_NIRCam} displays examples of best-fit reference models (which include all seven or eight chemical species) to NIRCam spectra.  When the contribution of clouds is removed from the model spectrum via post-processing, one sees that it is essentially identical to the reference model curve, implying that the clouds have no effect on the spectrum.  The effects of clouds on the NIRCam spectrum is negligible regardless of whether grey or non-grey clouds are assumed.  

For the NIRISS spectra, carbon dioxide is not detected regardless of the complexity of the temperature-pressure profile (Fig.~\ref{fig:Quest}).  The non-detection of carbon dioxide is consistent with the finding of \cite{Feinstein2023Natur.614..670F}.  Retrievals with grey clouds have upper limits that are optically thick, whereas those with non-grey clouds have optical depths of about unity (at 1 $\mu$m).  The retrieved water abundances are somewhat consistent between the eight different retrievals (four types of temperature-pressure profiles and grey versus non-grey cloud models).

\begin{figure*}[ht]
    \centering
    \includegraphics[width=0.95\linewidth]{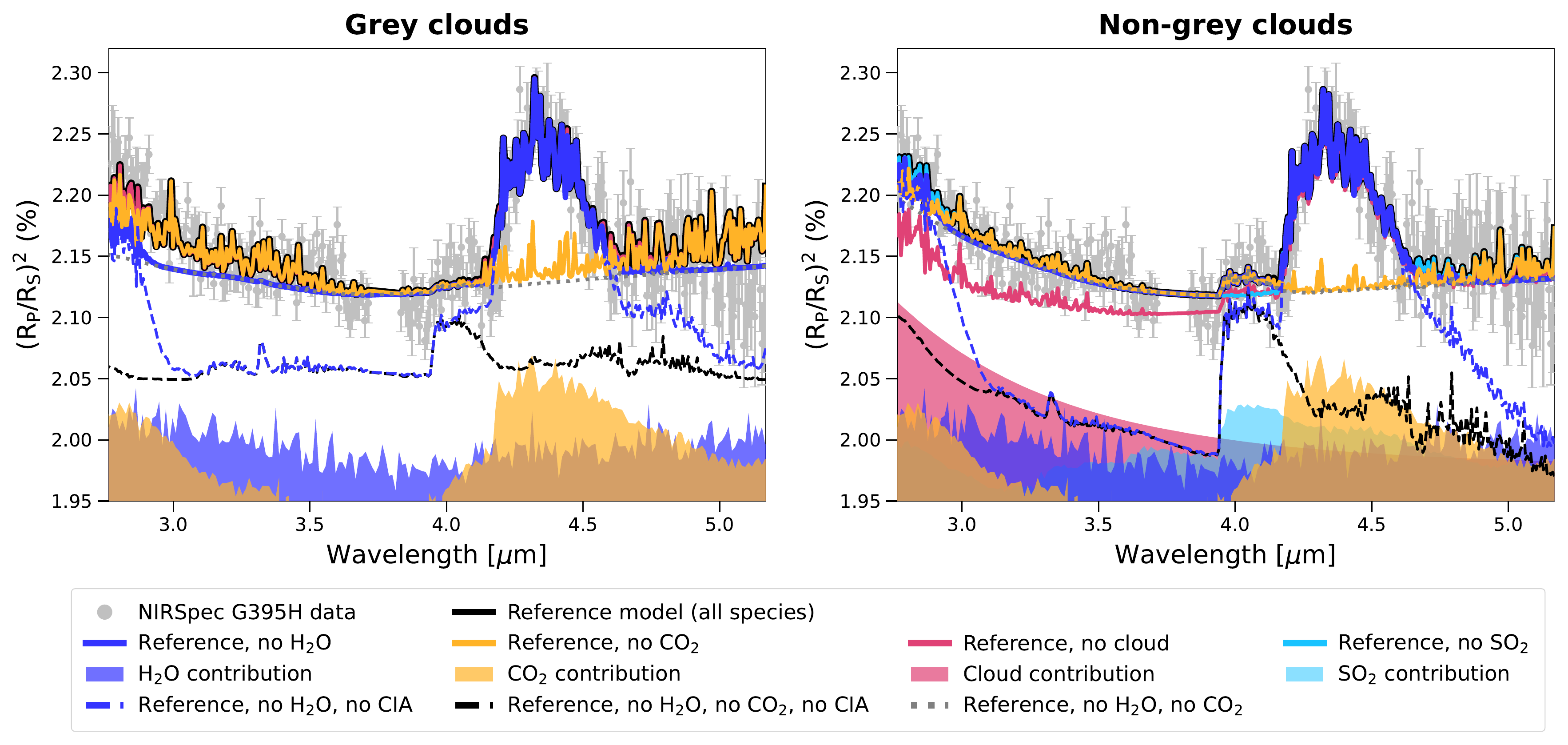}
    \caption{Fitting the model containing all chemical species (termed the ``reference model'') to the NIRSpec G395H spectrum (grey data points) previously published by \cite{Alderson2023Natur.614..664A}.  Corresponding to our findings in Fig.~\ref{fig:Quest}, the retrieval assuming grey clouds (left panel) uses an isothermal profile while the retrieval assuming non-grey clouds (right panel) uses a two-parameter temperature-pressure profile.  In each panel, post-processing is performed to isolate the contribution of water, carbon dioxide, sulphur dioxide, clouds, and collision-induced absorption (CIA) due to hydrogen and helium. The natural logarithm of the Bayes factor between the retrievals with grey and non-grey clouds is 33.8, with the non-grey ones being strongly preferred.}
    \label{fig:postprocessed_G395H}
\end{figure*}

For the NIRSpec G395H spectra, a major outcome of the current study is that the retrieved water abundances depend on whether grey or non-grey cloud models are used.  Fig.~\ref{fig:Quest} shows that the grey clouds are optically thin regardless of the temperature-pressure profile used, while the non-grey clouds are optically thick.  Fig.~\ref{fig:postprocessed_G395H} corroborates this finding, where the model spectrum with grey clouds removed (via post-processing) is identical to the reference model curve (with all chemical species and grey clouds included).  By contrast, the non-grey clouds have a non-negligible effect.  The spectral continuum associated with the non-grey cloud model compensates for the spectral continuum associated with the water line wings, altering the retrieved abundance of water.  

Water abundances derived from the NIRSpec PRISM spectrum are somewhat robust to the choice of cloud model, at least to within an order of magnitude (Fig.~\ref{fig:Quest}).  Fig. \ref{fig:postprocessed_PRISM} shows that the effects of clouds (whether grey or non-grey) are limited to $\lesssim 2.5$ $\mu$m when fitting model spectra to the NIRSpec PRISM spectrum.  This is consistent with our finding that the NIRCam spectrum is well fitted by models with optically thin clouds.  The compensation of the spectral continuum by the non-grey-cloud model at $\sim$3 $\mu$m, which is seen for the NIRSpec G395H spectrum (Fig.~\ref{fig:postprocessed_G395H}), is not present for the NIRSpec PRISM spectrum, suggesting that this compensation is spurious. It further suggests that having a broader wavelength coverage is more important than higher spectral resolution if the goal is to better understand the effects of clouds on the transmission spectrum.  The effects of clouds on the NIRSpec PRISM spectrum has major implications for the detection or non-detection of chemical species, as we will explore later.

\subsubsection{Detection or non-detection of chemical species}

\begin{figure*}
\centering
\includegraphics[width=0.85\textwidth]{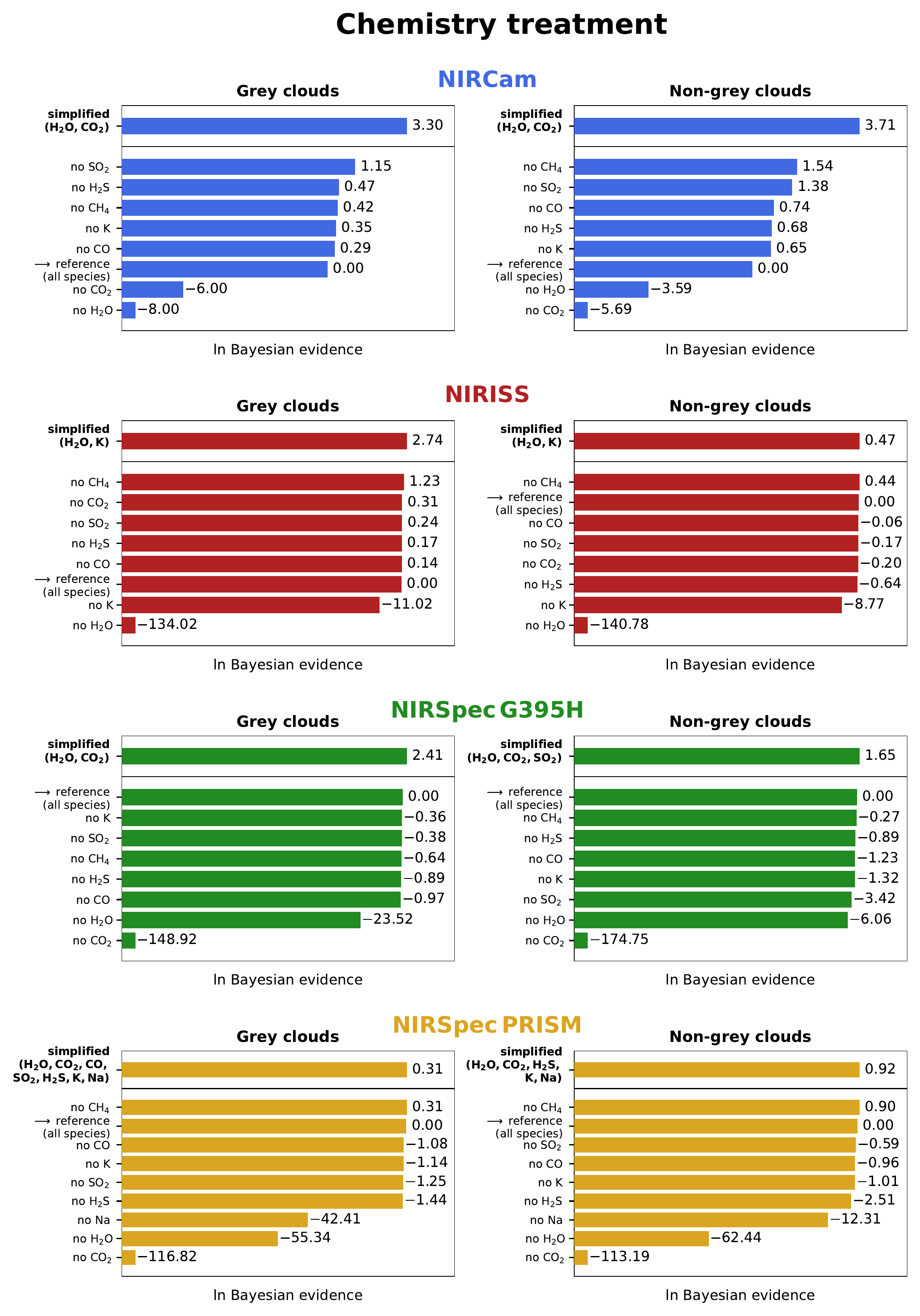}
\caption{Bayesian model comparison between models that include different subsets of chemical species.  Starting with the reference model (which includes all chemical species), a selected atom or molecule is excluded.  If the Bayesian evidence decreases significantly (such that the logarithm of the Bayes factor is much greater than unity), then this chemical species is considered part of the ``simplified model'' (which contains the minimal set of species required to fit the data).  All other chemical species that produce increases in the Bayesian evidence when they are excluded from the retrieval are thus excluded from the simplified model.  In each of the panels, the natural logarithm of the Bayes factor (relative to the reference model) is displayed.  Retrievals assuming grey (left column) versus non-grey (right panel) clouds are considered separately.  The sophistication of the temperature-pressure profile used is based on the findings of Fig.~\ref{fig:Quest}.}
\label{fig:bayes_chemistry}
\end{figure*}

To determine if a specific chemical species is detected, we ran a second suite of retrievals using the reference model (which includes all chemical species) as a starting point.  Each of the seven (or eight for PRISM retrievals) chemical species is then excluded, in turn, in separate retrievals.  The logarithm of the Bayes factor $\ln{\cal B}$, relative to the reference model, is recorded.  Fig.~\ref{fig:bayes_chemistry} shows the outcome of this suite of retrievals for all four JWST instrument modes.  If the exclusion of an atom or molecule results in a decrease in the Bayesian evidence, relative to the reference model, \textit{and} yields $\ln{\cal B}>1$, then we deem it part of the minimal set of chemical species required to explain the data.  By contrast, if the exclusion of an atom or molecule results in an increase in the Bayesian evidence (regardless of the Bayes factor value), then we exclude it from this ``simplified model''.  We consider the chemical species that are part of the simplified model to be detected in the data.

Using this approach, Fig.~\ref{fig:bayes_chemistry} shows that only \ch{H2O} and \ch{CO2} are required to fit the NIRCam spectrum.  Our derived \ch{CO2} abundances from the grey-cloud and non-grey-cloud retrievals are consistent with each other: $\log X_{\rm CO_2} = -3.34_{-0.53}^{+0.44}$ and $-3.30_{-0.52}^{+0.43}$, respectively.  For completeness, Figs.~\ref{fig:NIRCam_Corner_fav} and ~\ref{fig:NIRCam_Corner_unfav} display the full set of posterior distributions of parameters.  We checked that the optical depth of the grey cloud is much less than unity (Fig.~\ref{fig:NIRCam_Corner_fav}).  We also checked that the range of optical depths for the non-grey cloud, across the NIRCam range of wavelengths, is much less than unity (median values $\sim$10$^{-5}$--$10^{-4}$; not shown).

For the NIRISS spectrum, only \ch{H2O} and \ch{K} are required to explain the data.  However, the derived logarithm of the potassium abundances are inconsistent, because of the effect of clouds: $-7.50_{-0.40}^{+0.41}$ (grey clouds) versus $-5.59_{-0.51}^{+0.46}$ (non-grey clouds). Again for completeness, Figs.~\ref{fig:NIRISS_Corner_fav} and ~\ref{fig:NIRISS_Corner_unfav} display the full set of posterior distributions of parameters.  The optical depth of the grey cloud is well above unity (Fig.~\ref{fig:NIRISS_Corner_unfav}).  Across the NIRISS range of wavelengths, the optical depth of the non-grey cloud ranges from $\sim$10$^{-2}$--10 (not shown).

For the NIRSpec G395H spectrum, the simplified model includes \ch{H2O} and \ch{CO2}.  However, when a non-grey cloud model is used, one has to additionally include \ch{SO2}.  Furthermore, the derived logarithm of the \ch{H2O} abundances are noticeably different depending on whether the grey or non-grey cloud model is used: $-5.28_{-0.36}^{+0.37}$ (grey cloud) versus $-6.35_{-0.33}^{+0.34}$ (non-grey cloud).  By contrast, we have for the logarithm of the \ch{CO2} abundances: $-6.46_{-0.37}^{+0.44}$ (grey cloud) versus $-6.49_{-0.28}^{+0.34}$ (non-grey cloud). The full set of posterior distributions of parameters are displayed in Figs.~\ref{fig:G395H_Corner_fav} and ~\ref{fig:G395H_Corner_unfav}.  The grey cloud is optically thin (Fig.~\ref{fig:G395H_Corner_unfav}), whereas the non-grey cloud is optically thick across part of the NIRSpec G395H range of wavelengths.

\begin{figure*}[ht]
    \centering
    \includegraphics[width=\linewidth]{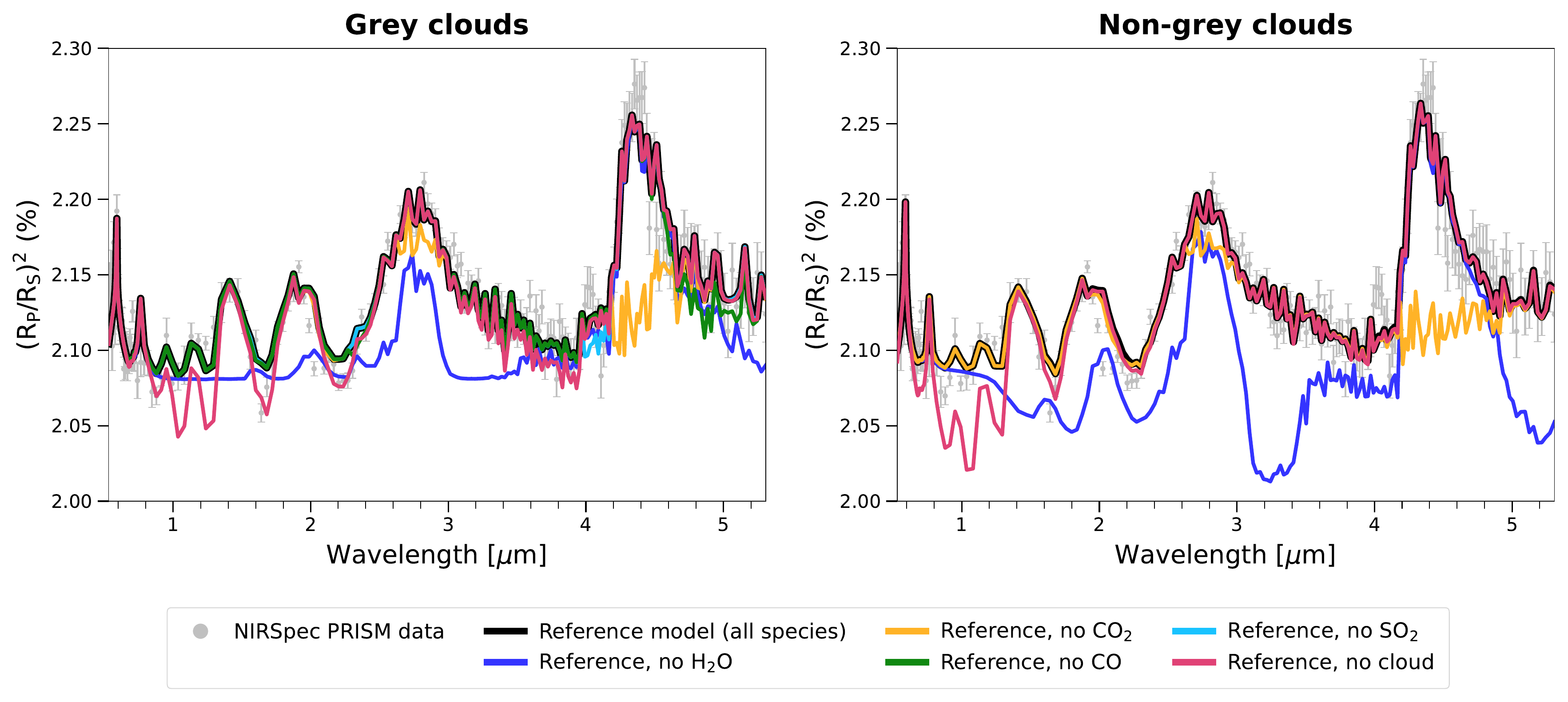}
    \caption{Fitting the model containing all chemical species (termed the ``reference model'') to the NIRSpec PRISM spectrum (grey data points) previously published by \cite{Rustamkulov2023Natur.614..659R}.  Corresponding to our findings in Fig.~\ref{fig:Quest}, the retrievals assuming grey clouds (left panel) and non-grey clouds (right panel) both use isothermal temperature-pressure profiles.   In each panel, post-processing is performed to isolate the contribution of water, carbon dioxide, carbon monoxide, sulphur dioxide, and clouds. The natural logarithm of the Bayes factor between this pair of retrievals is 0.61, implying that Bayesian model comparison has no preference for grey versus non-grey clouds.}
    \label{fig:postprocessed_PRISM}
\end{figure*}

For NIRSpec PRISM, methane is not clearly detected (based on Bayesian model comparison) regardless of whether grey or non-grey clouds are used, consistent with the findings of \cite{Rustamkulov2023Natur.614..659R}.  \ch{CO} and \ch{SO2} are only clearly detected when grey clouds are assumed.  When non-grey clouds are assumed, the best-fit water opacity is altered, which in turn compensates for the spectral features of \ch{CO} and \ch{SO2} (Fig. \ref{fig:postprocessed_PRISM}).  In other words, the compensation of the spectral features by the non-grey cloud model is indirect and occurs via the water spectral features.  Figs.~\ref{fig:PRISM_Corner_fav} and ~\ref{fig:PRISM_Corner_unfav} display the full set of posterior distributions of parameters. The grey cloud is optically thin, while the non-grey cloud has a range of optical depths $\sim$10$^{-1}$--$10^{3}$ across the PRISM range of wavelengths.

\subsubsection{Instrument-specific best fit spectra}

\begin{figure*}[ht]
    \centering
    \includegraphics[width=\textwidth]{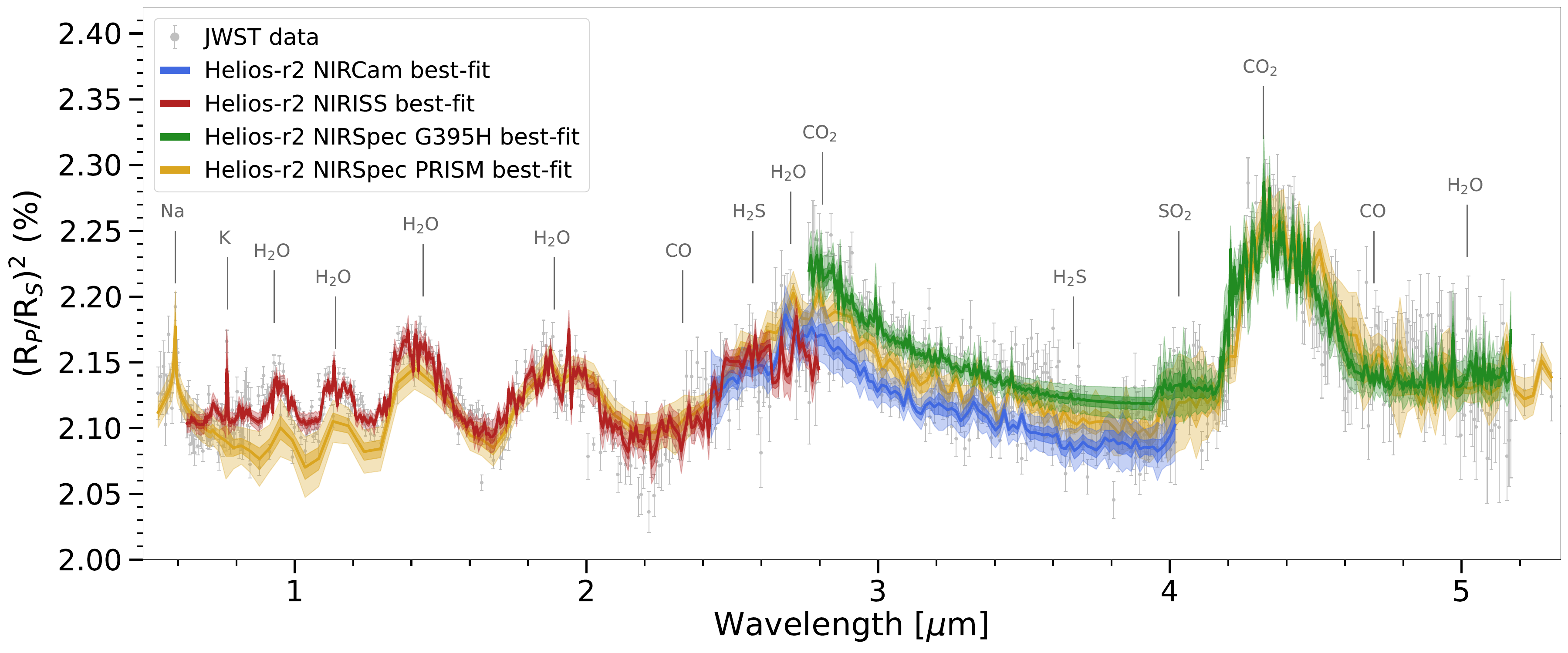}
    \vspace{-0.5cm}
    \caption{Best-fit models to the various measured spectra (grey data points) from our nested sampling retrievals.  The shaded regions associated with the model curves show fit uncertainties of 2 and 5 standard deviations.  Spectral features corresponding to various chemical species are labelled.  The best-fit models to NIRCam and NIRSpec PRISM spectra assume a grey-cloud model, while those to NIRISS and NIRSpec G395H assume a non-grey-cloud model.  Full posterior distributions of parameters corresponding to these fits are provided in Figs. \ref{fig:NIRCam_Corner_fav} to \ref{fig:PRISM_Corner_fav}.}
    \label{fig:HELIOS_BF}
\end{figure*}

The best-fit spectra from our suite of nested-sampling retrievals are shown in Fig.~\ref{fig:HELIOS_BF}.  For each spectrum, we choose between the grey versus non-grey cloud model using the logarithm of the Bayes factor. For the NIRCam spectrum, the Bayesian evidence is higher for the grey-cloud model, but the logarithm of the Bayes factor is only $\ln{\cal B}=0.41$, implying that the non-grey cloud model fits the data equally well.  As previously explained, this is because both the grey and non-grey clouds are optically thin and the model is effectively cloud-free. For the NIRISS spectrum, Bayesian model comparison strongly favours the non-grey cloud model with $\ln{\cal B}=11$.  

That the non-grey cloud model is preferred for the NIRSpec G395H spectrum ($\ln{\cal B}=34$), despite its spectral continuum spuriously compensating for the retrieved water abundances, is a cautionary tale for the uncritical acceptance of Bayesian model comparison implemented via nested sampling.  In particular, this study has shown that parametric cloud models of capable of mimicking gas absorption features if they have multiple degrees of freedom.

For the NIRSpec PRISM spectrum, Bayesian model comparison is agnostic between the grey and non-grey cloud models ($\ln{\cal B}=0.61$).  But as Fig. \ref{fig:bayes_chemistry} and \ref{fig:postprocessed_PRISM} already demonstrate, \ch{CO} and \ch{SO2} are detected only when the grey-cloud model is assumed. 

The full sets of posterior distributions are provided in the appendix (Figs.~\ref{fig:NIRCam_Corner_fav} to \ref{fig:PRISM_Corner_unfav}), while the retrieved values of the parameters are stated in Table~\ref{tab:heliosr2_posteriors}.  Generally, the retrieved values of $\log{g}$, $R_p$, and $R_s$ are consistent with the assumed prior distributions to within one standard deviation.  Since these assumed prior distributions are based on the measurements of \citet{Mancini+18}, it implies that the planetary radius, mass and gravity are self-consistent with one another.  An exception is the inferred planetary gravity from the retrieval performed on the PRISM spectrum assuming non-grey clouds: $\log{g} = 2.77 \pm 0.02$, which is discrepant with the assumed prior value of $2.63 \pm 0.05$. We speculate that the enhanced gravity comes from the retrieval seeking a diminished pressure scale height to fit the spectral features. However, it is unclear why this occurs only for this particular case, and is possibly a combination of effects from multiple parameters.


\subsubsection{Elemental abundances}

\begin{figure}[ht]
    \centering
    \includegraphics[width=0.95\columnwidth]{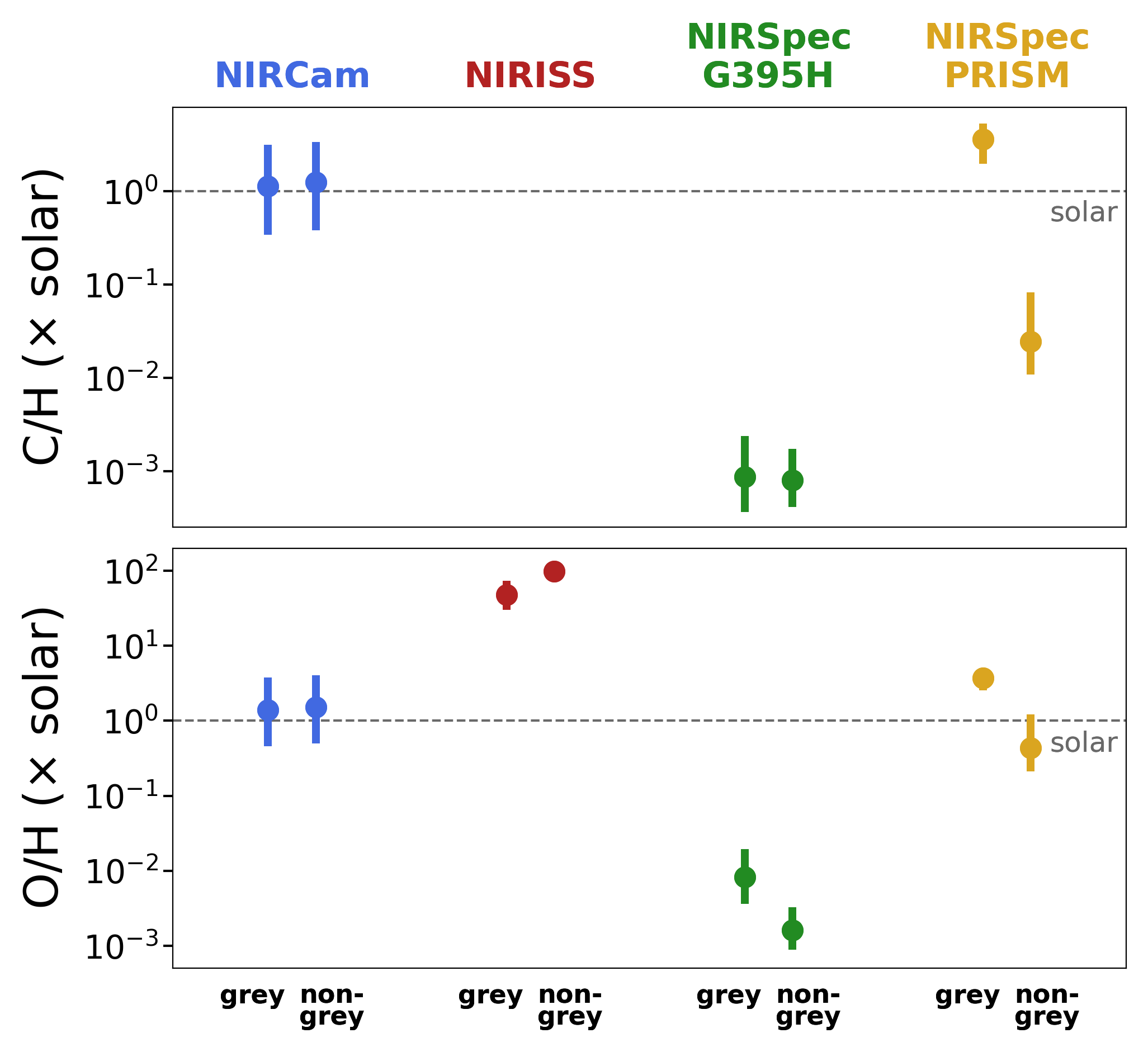}
    \vspace{-0.2cm}
    \caption{Retrieved elemental abundances (relative to their solar values) from the simplified models (minimal set of chemical species needed to fit data).  The solar elemental abundances assumed are $\mbox{C/H} = 2.95 \times 10^{-4}$ and $\mbox{O/H} = 5.37\times 10^{-4}$ \citep{Asplund2009ARA&A..47..481A}.  Separate retrievals are performed assuming grey versus non-grey clouds.  Displayed are the median values of the posterior distributions and their 1-$\sigma$ uncertainties. There is no C/H entry for NIRISS, because no carbon-bearing species are detected from that spectrum.}
    \label{fig:CtoH_OtoH_ratios}
    \vspace{-0.4cm}
\end{figure}

Figure \ref{fig:CtoH_OtoH_ratios} shows the retrieved elemental abundances of carbon (C/H) and oxygen (O/H), which are sometimes referred to as the ``metallicity''.  The retrieved C/H and O/H values from NIRCam are consistent with each other between the grey and non-grey cloud models, because the clouds are optically thin.  C/H is not retrieved from NIRISS, because no carbon-bearing molecule is detected.  For NIRSpec G395H, the retrieved C/H values are consistent with each other between the grey and non-grey cloud models, while the O/H values are discrepant by about an order of magnitude.  The retrieved C/H values from NIRSpec PRISM are somewhat discrepant between the grey- and non-grey-cloud retrievals by a factor $\sim$100.  Overall, the inferred C/H and O/H values vary by factors $\sim$10$^3$ and $\sim$10$^5$, respectively, between the four different instruments, implying that deriving metallicities from a single instrument mode should be done with caution.

Except for O/H derived from NIRISS, the retrieved C/H and O/H values are sub-solar to solar (relative to the \citealt{Asplund2009ARA&A..47..481A} values).  However, these values cannot be directly compared to those reported by the ERS papers \citep{Alderson2023Natur.614..664A, Ahrer2023Natur.614..653A, Feinstein2023Natur.614..670F, Rustamkulov2023Natur.614..659R}, because these were derived using model grids that assume chemical equilibrium and reported as the sum of elemental abundances.

\subsection{Random forest retrievals}

\begin{figure*}[ht]
    \centering
    \includegraphics[width=\textwidth]{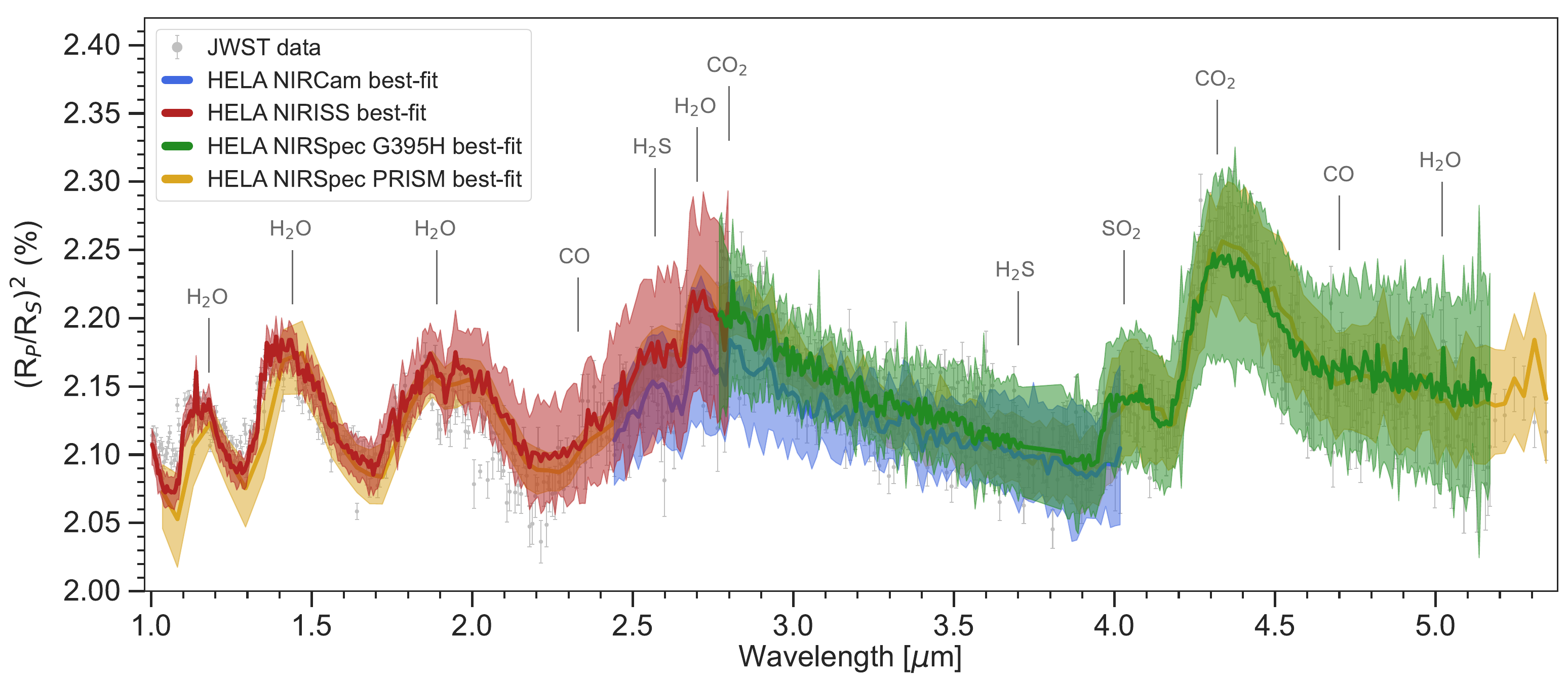}
    \vspace{-0.3cm}
    \caption{Best-fit models to the various measured spectra (grey data points) from our random forest retrievals trained on the model grid of \cite{Crossfield2023ApJ...952L..18C}.  The shaded regions associated with the model curves show fit uncertainties of 1 standard deviation.  Spectral features corresponding to various chemical species are labelled.}
    \label{fig:HELA_BF}
\end{figure*}

\begin{figure}[ht]
    \centering
    \includegraphics[width=\columnwidth]{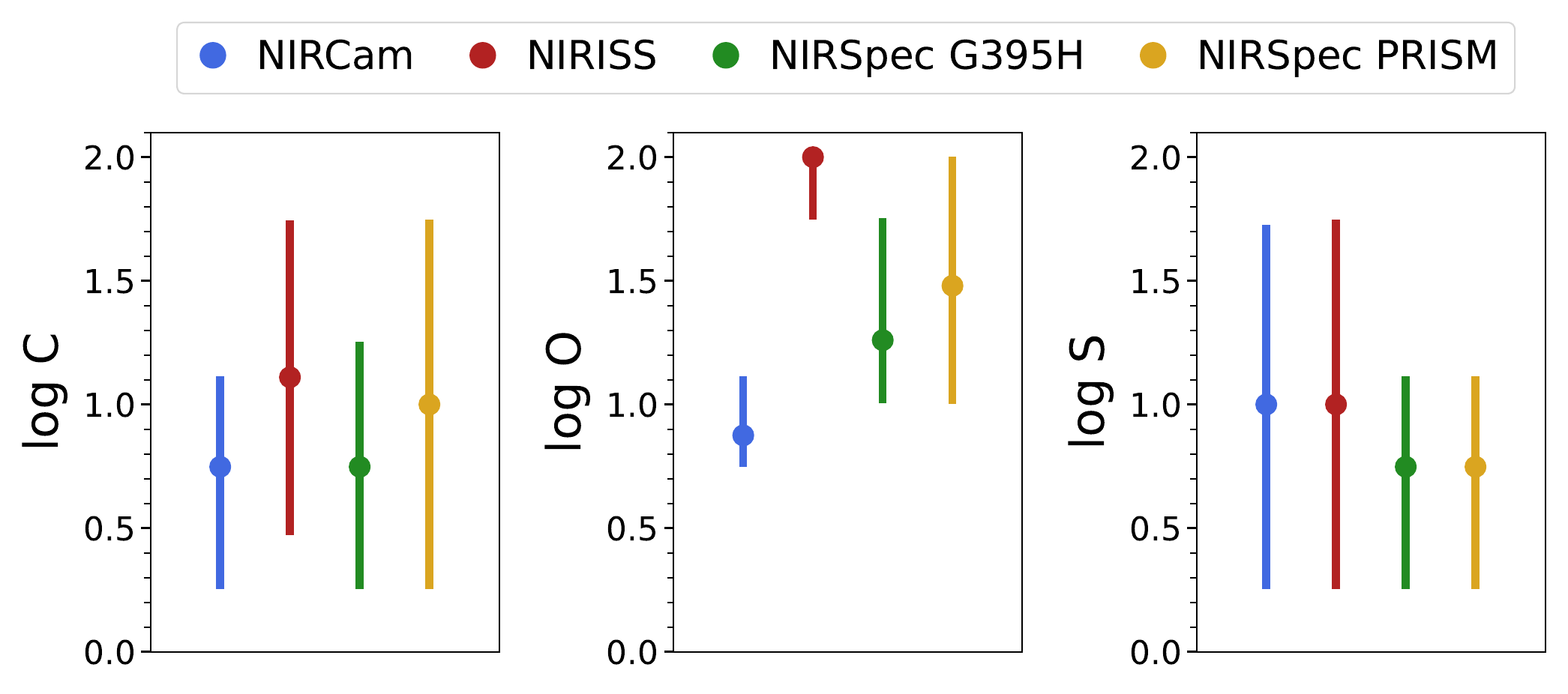}
    \vspace{-0.3cm}
    \caption{Median values (and their 1-$\sigma$ uncertainties) of elemental abundances inferred from the median values of the posterior distributions associated with our random forest retrievals.  Separate entries for each JWST instrument mode are given.  The elemental abundances are given in terms of their solar values.}
    \label{fig:HELA_posteriors}
\end{figure}

In addition to the nested-sampling retrievals, we performed another suite of random-forest retrievals using the \cite{Crossfield2023ApJ...952L..18C} model grid as the training set.  The best-fit spectra and inferred \ch{C}, \ch{O}, and \ch{S} elemental abundances are displayed in Figs.~\ref{fig:HELA_BF} and \ref{fig:HELA_posteriors}. 
The retrieved values of the parameters are stated in Table~\ref{tab:hela_posteriors}, while the full sets of posterior distributions are provided in the appendix (Figs.~\ref{fig:HELA_posteriors_NIRCam} to \ref{fig:HELA_posteriors_PRISM}).

We note that we exclude all data points below 1.0 $\mu$m for the NIRISS and NIRSpec PRISM spectra, as the model grid only includes wavelengths above that value.  All four retrievals consistently yield super-solar metallicities, albeit with generous uncertainties, consistent with the findings of the ERS team \citep{Alderson2023Natur.614..664A, Ahrer2023Natur.614..653A, Feinstein2023Natur.614..670F, Rustamkulov2023Natur.614..659R}.  

The random forest retrievals are able to report C/H values from NIRISS spectra, probably because carbon-carrying molecules are assumed to be present.  The \cite{Crossfield2023ApJ...952L..18C} model grid extends only to elemental abundances of $100\times$ solar.  Only the O/H value associated with NIRISS approaches that boundary.  

Previous studies have shown that the random forest method tends to be conservative with uncertainty estimation compared to the nested sampling method \citep{Marquez-Neila2018NatAs...2..719M, Oreshenko2020AJ....159....6O, Fisher2020AJ....159..192F, Lueber2023ApJ...954...22L}.  These larger uncertainties are also reflected in the best-fit spectra (Fig.~\ref{fig:HELA_BF}).  Another reason for the large model uncertainties is the sparseness of the training set of models.

\subsection{Model dependency of C/O ratios}

Given the retrieved molecular abundances, the C/O ratio is constructed using
\begin{equation}
\mbox{C/O} = \frac{X_{\ch{CO}} + X_{\ch{CO2}} + X_{\ch{CH4}}}{X_{\ch{CO}} + 2X_{\ch{CO2}} + X_{\ch{H2O}} + 2X_{\ch{SO2}}},
\end{equation}
where $X_i$ is the volume mixing ratio (relative abundance by number) of species $i$.

Earlier, it was demonstrated that the retrieved water abundances from the NIRSpec G395H spectrum are model-dependent.  Therefore, we refrain from constructing C/O ratios using retrieved molecular abundances from nested-sampling retrievals performed on the NIRSpec G395H spectrum.  Since the nested-sampling retrievals do not require carbon-bearing species to fit the NIRISS spectrum, we do not construct C/O ratios from it as well.

By contrast, water abundances retrieved from the NIRSpec PRISM spectrum appears to be robust to the choice of cloud model.  For the nested sampling retrievals, five of these models are associated with the logarithm of the Bayes factor being less than unity.  Therefore, we construct C/O ratios from the retrieved chemical abundances.  For the NIRCam spectrum, we construct C/O ratios only from the pair of models with 3-parameter temperature-pressure profiles (based on Bayesian model comparison).

\begin{figure}[ht]
    \centering
    \includegraphics[width=\columnwidth]{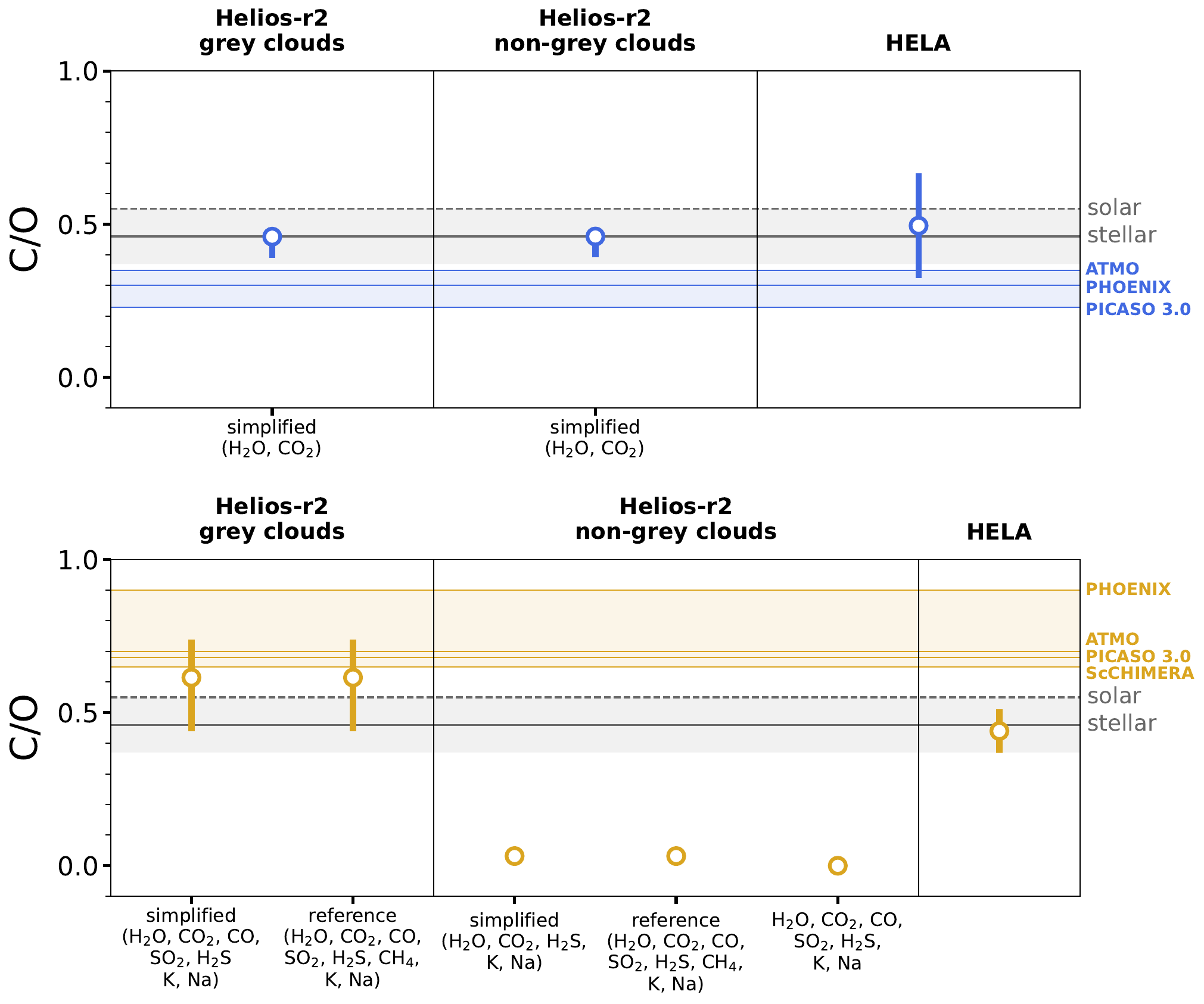}
    \vspace{-0.4cm}
    \caption{Median values of the carbon-to-oxygen ratio (C/O) inferred from both our nested sampling and random forest retrievals.  Only retrievals performed on NIRCam (top panel) and NIRSpec PRISM (bottom panel) spectra are considered (see text for more details).  Only models with the logarithm of the Bayes factor being less than unity (relative to the model with the highest Bayesian evidence) are shown.  The three horizontal blue lines are C/O ratios taken from \cite{Ahrer2023Natur.614..653A} for NIRCam, while the four horizontal yellow lines are those taken from \cite{Rustamkulov2023Natur.614..659R} for NIRSpec PRISM. The dashed grey line indicates the solar value of 0.55 \citep{Asplund2009ARA&A..47..481A}, while the solid grey line represents the stellar value of 0.46 $\pm$ 0.09 \citep{Polanski2022RNAAS...6..155P}.}
    \label{fig:CtoO_ratios}
    \vspace{-0.3cm}
\end{figure}

\begin{figure}[ht]
    \centering
    \includegraphics[width=0.95\columnwidth]{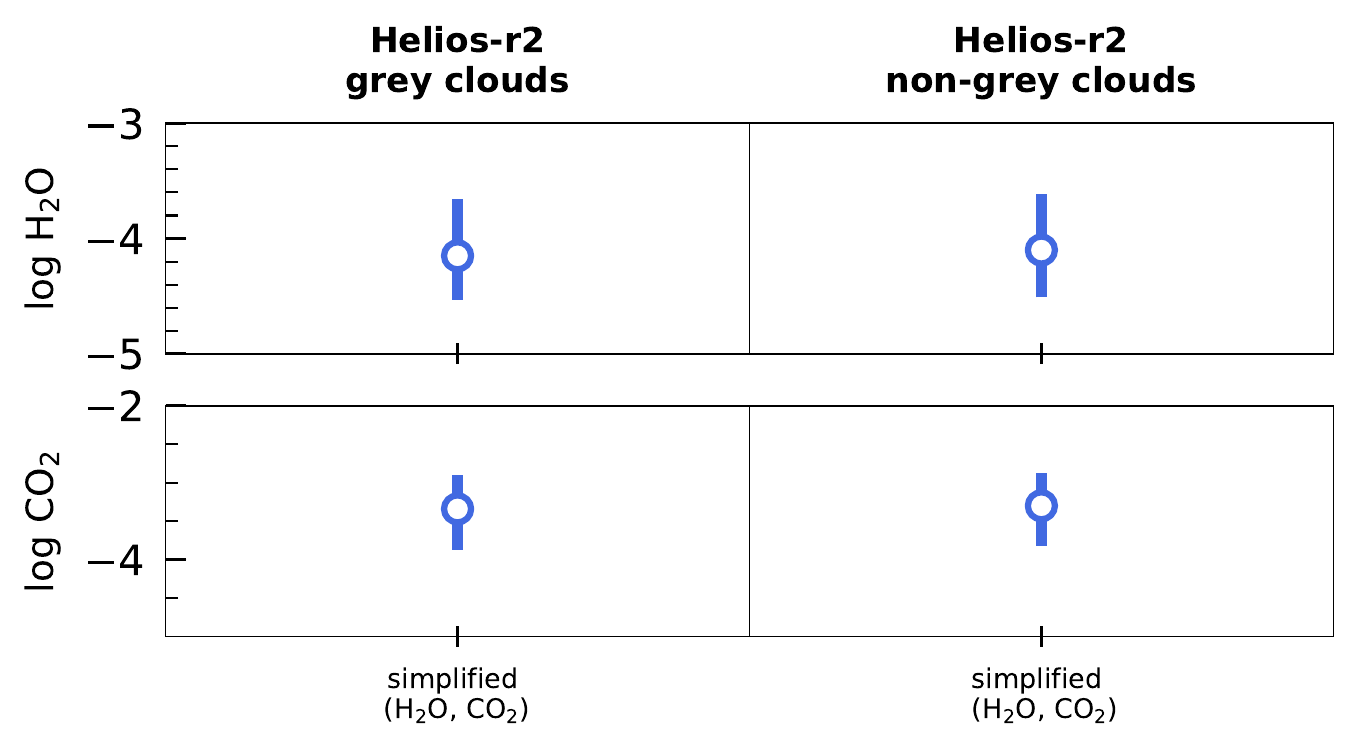}
    \includegraphics[width=0.98\columnwidth]{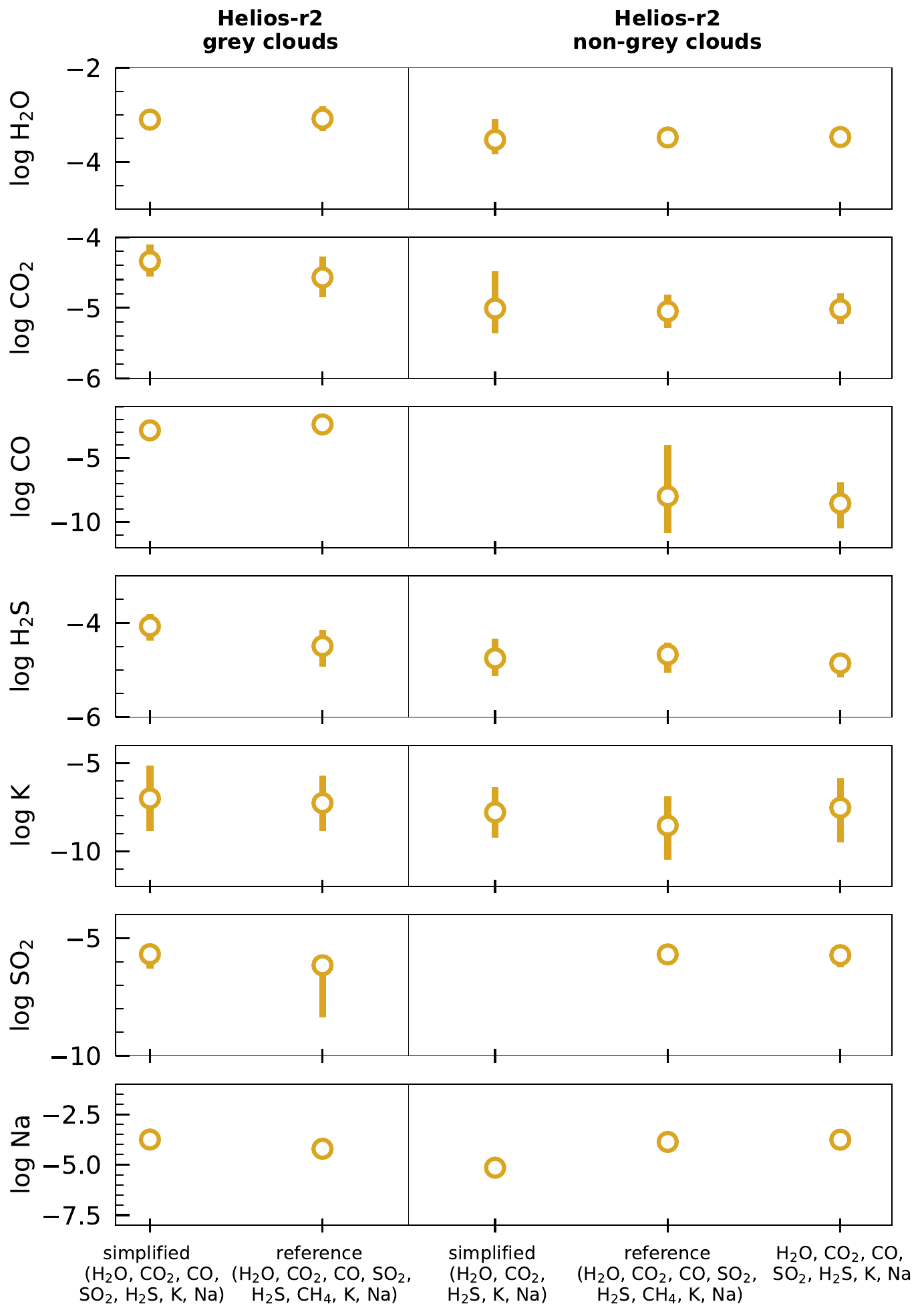}
    \vspace{-0.1cm}
    \caption{Retrieved chemical abundances (volume mixing ratios) corresponding to the models shown in Figure \ref{fig:CtoO_ratios}. Displayed are the median values of the posterior distributions and their 1-$\sigma$ uncertainties.}
    \label{fig:Posteriors_PRISM}
    \vspace{-0.3cm}
\end{figure}

Figure \ref{fig:CtoO_ratios} shows that the C/O ratios derived from the NIRCam spectrum are consistent between the retrievals with grey versus non-grey clouds, because both of these clouds are optically thin, i.e. the retrievals are effectively cloudfree.  The derived values are consistent with the stellar C/O value of 0.46 $\pm$ 0.09 \citep{Polanski2022RNAAS...6..155P}, but inconsistent with substellar (and sub-solar) values derived by the ERS team \citep{Ahrer2023Natur.614..653A}.  Figure \ref{fig:Posteriors_PRISM} demonstrates that the derived water and carbon dioxide abundances are consistent between the grey-cloud and non-grey-cloud retrievals.  The C/O ratio derived from the random forest retrievals, using the \cite{Crossfield2023ApJ...952L..18C} model grid, has larger uncertainties but is also consistent with the stellar C/O.  

However, the C/O ratios derived from the NIRSpec PRISM spectrum depend on whether grey or non-grey clouds are assumed (Figure \ref{fig:CtoO_ratios}).  The ones inferred from grey-cloud retrievals are super-stellar (and super-solar), whereas the ones inferred from non-grey cloud models are sub-stellar (and sub-solar).  At face value, this finding contradicts our earlier claim that the retrieved PRISM water and carbon dioxide abundances are robust.  However, Figure \ref{fig:Posteriors_PRISM} demonstrates that, while the model dependency is diminished (compared to those associated with the NIRSpec G395H spectrum) these retrieved abundances still differ by about half an order of magnitude between grey-cloud and non-grey-cloud retrievals.  This variation is sufficient to render the retrieval C/O ratios model-dependent.

Unfortunately, Bayesian model comparison does not offer a way forward, because the logarithm of the Bayes factor is less than unity ($\ln{\cal B}=0.61$) when comparing the grey-cloud and non-grey-cloud retrievals associated with NIRSpec PRISM. Generally, accurate and precise C/O ratios are challenging to obtain, because variations in the models produce small changes in the logarithm of the chemical abundances, which lead to large changes in the C/O ratio.

\section{Summary and discussion}
\label{sect: discussion}

\subsection{Comparison to previous work}

Since WASP-39b is a benchmark object for JWST exoplanet science, due to its selection for being part of the ERS program, a detailed comparison with previous studies by the ERS team is indispensable.  

\subsubsection{NIRCam spectrum}

\cite{Ahrer2023Natur.614..653A} previously reported the detection of water and an upper limit on the abundance of methane from the NIRCam spectrum.  Our Bayesian model comparison analysis (Figure \ref{fig:bayes_chemistry}) is consistent with the detection of water and non-detection of methane.  \cite{Ahrer2023Natur.614..653A} reported how the ``prominent carbon dioxide feature at 2.8 micrometres is largely masked by water''.  Based on the Bayesian evidence, our nested sampling retrievals require \ch{H2O} and \ch{CO2} to fit the NIRCam spectrum.

\cite{Ahrer2023Natur.614..653A} inferred ``atmospheric metallicities of 1--100 times solar and a substellar C/O ratio''.  Our inferred C/O ratios, from both nested-sampling and random forest retrievals, are consistent with the stellar value (Fig. \ref{fig:CtoO_ratios}).  The source of this discrepancy remains unclear.

These metallicities are consistent with the elemental abundances inferred using our random forest retrievals (Fig. \ref{fig:HELA_posteriors}).  Our nested-sampling retrievals infer C/H and O/H values that are consistent with being solar (Fig. \ref{fig:CtoH_OtoH_ratios}).  

A significant difference between our analysis and that of \cite{Ahrer2023Natur.614..653A} is whether clouds are needed to fit the NIRCam spectrum.  Figure 3 of that study demonstrates clearly that the cloud model implemented by \cite{Ahrer2023Natur.614..653A} provides almost the entire spectral continuum both blueward and redward of the peak of the 2.8 $\mu$m \ch{CO2} feature.

In Fig. \ref{fig:postprocessed_NIRCam}, we demonstrate that water and CIA (associated with hydrogen and helium) provide the opacity sources for the spectral continuum.  When the contributions of \ch{CO2} and \ch{H2O} are removed via post-processing, a featureless continuum remains that is due to CIA.  These properties explain why the clouds in our NIRCam retrievals are optically thin (transparent), regardless of whether they are grey or non-grey.  It also explains why our nested-sampling retrievals are able to robustly extract \ch{CO2} abundances, despite the water opacity partially obscuring the carbon dioxide opacity in the same wavelength range.

\subsubsection{NIRISS spectrum}

\cite{Feinstein2023Natur.614..670F} previously reported the detection of water and potassium from the NIRISS spectrum.  This is confirmed by our Bayesian model comparison analysis (Figure \ref{fig:bayes_chemistry}), which requires only these two species to fit the NIRISS spectrum.  We are unable to verify the sub-solar C/O ratio reported by \cite{Feinstein2023Natur.614..670F} as our nested-sampling retrievals do not require carbon-bearing species to fit the NIRISS spectrum, and therefore we are unable to compute C/O ratios.  Our random forest retrievals, which are trained on the \cite{Crossfield2023ApJ...952L..18C} model grid, return a C/O ratio of $0.63 \pm 0.15$.

\subsubsection{NIRSpec G395H spectrum}

\cite{Guzman2020AJ....160...15G} previously suggested that the NIRSpec G395H mode is optimal for detecting multiple species of simple carbon-, oxygen-, and hydrogen-carrying molecules.  \cite{Alderson2023Natur.614..664A} confirmed this prediction by reporting the detection of carbon dioxide, water, and sulphur dioxide from the NIRSpec G395H spectrum.  Our Bayesian model comparison (Figure \ref{fig:bayes_chemistry}) confirms the detection of water and carbon dioxide.  However, whether sulphur dioxide is detected depends on whether a grey or non-grey cloud model is used, an issue we will investigate in the next subsection.  CIA associated with hydrogen and helium again provides much of the spectral continuum (Fig. \ref{fig:postprocessed_G395H}).


\subsubsection{NIRSpec PRISM spectrum}

\cite{Rustamkulov2023Natur.614..659R} previously reported the detection of water, carbon monoxide, carbon dioxide, sodium, and sulphur dioxide from the NIRSpec PRISM spectrum, as well as the non-detection of methane.  Our Bayesian model comparison (Fig.~\ref{fig:bayes_chemistry}) confirms these findings and also reports the detection of hydrogen sulphide.  However, whether sulphur dioxide is detected depends again on whether a grey or non-grey cloud model is used.  

Fig. \ref{fig:CtoO_ratios} demonstrates that the C/O ratios from our nested-sampling retrievals are consistent with the super-solar values reported by \cite{Rustamkulov2023Natur.614..659R} when grey-cloud models are assumed, but inconsistent with them when non-grey clouds are assumed.  Our random forest retrievals return a C/O ratio that is consistent with the stellar value.  Bayesian model comparison does not offer us a way out of these discrepancies, because the logarithm of the Bayes factor is less than unity between the grey-cloud and non-grey-cloud nested sampling retrievals.

\subsection{Is sulphur dioxide detected?}

Earlier, we demonstrated that the use of a non-grey cloud model produces a spectral continuum at $\sim$3 $\mu$m that compensates for the spectral continuum associated with water in the NIRSpec G395H spectrum (Fig. \ref{fig:postprocessed_G395H}).  The same figure shows that the presence of the non-grey cloud diminishes the \ch{H2O} abundance enough that it allows for the spectral feature at 4.0 $\mu$m to be fitted by \ch{SO2}.

When the same non-grey cloud model is fitted to the NIRSpec PRISM spectrum, it does not compensate for the spectral continuum over the same wavelength range (Fig. \ref{fig:postprocessed_PRISM}).  Given the broader wavelength coverage of PRISM, it suggests that this compensation by the non-grey cloud model, for the fit to the G395H spectrum, is spurious.

The non-grey cloud model influences the required water abundance needed to fit the NIRSpec PRISM spectrum, which in turn leaves no room for the spectral feature at 4.0 $\mu$m to be fitted by \ch{SO2}. sulphur dioxide is needed for the model fit when grey clouds are assumed (which is also paired with an isothermal profile).  Physically, grey clouds are possible when the cloud particle radius exceeds the longest wavelength covered by the spectrum (divided by $2\pi$), i.e. $5.5/2\pi \sim 1$ $\mu$m, which is a plausible particle size.  There is no preference for either the grey or non-grey cloud model ($\ln{\cal B}=0.61$) when all chemical species are included in the retrieval (the so-called ``reference model''), as well as when only the minimal set of molecules are included (the so-called ``simplified model'').

While the spectral feature at 4.0 $\mu$m is real, its interpretation (and the detection of \ch{SO2}) appears to be model-dependent.  Bayesian model comparison does not offer us a way out of this conundrum. Generally, we advocate for the robustness of detections to be checked using spectra measured by multiple instruments subjected to different data reduction techniques and interpreted using several retrieval codes.

\subsection{The normalization degeneracy of transmission spectra}

\cite{BennekeSeager2012ApJ...753..100B} and \cite{Griffith2014RSPTA.37230086G} previously identified a degeneracy between the absolute value of a transmission spectrum and its encoded molecular abundances.  \cite{Heng2017MNRAS.470.2972H} showed that this ``normalization degeneracy'' may be explained analytically and is a three-way degeneracy between an arbitrary reference transit radius, its corresponding reference pressure and the total cross section or opacity of the atoms and molecules in the atmosphere.  \cite{FisherHeng18} suggested that the shape of spectral features and the continuum could encode enough pressure-dependent information to break this degeneracy.

Figs.~\ref{fig:NIRCam_Corner_fav} to ~\ref{fig:PRISM_Corner_unfav} demonstrate that the joint posterior distributions between $R_p$ (which is essentially the normalization of each transmission spectrum as it is the transit radius at 10 bar) and the abundances of atoms and molecules are not degenerate, meaning that these abundances do not depend on the normalization.  The width of the posterior distribution of $R_p$ is determined by the degeneracy with the stellar radius.  A different choice of the stellar radius would simply produce a different fit for $R_p$ without influencing the retrieved chemical abundances.  While the normalization of transmission spectra should remain a fitting parameter in retrievals, we conclude that it should not be degenerate with the retrieved chemical abundances for high-quality JWST spectra.

The degeneracies are between the different molecular abundances themselves.  For example, the degeneracy between \ch{CO2} and \ch{H2O} abundances in the interpretation of the NIRCam spectrum corresponds to the obscuration of the opacity of the former by that of the later.  

\subsection{Summary}

In the current study, we have investigated the information content of the JWST spectra of WASP-39b measured by four different instruments.  Our main findings are:
\begin{itemize}

\item The complexity of the temperature-pressure profile required to fit the data depends on the instrument mode used (and the wavelength coverage).

\item The minimum set of atoms and molecules required to fit a spectrum depends on the instrument mode used.  


\item Using a non-grey cloud model to fit the NIRSpec G395H spectrum results in a spectral continuum that spuriously compensates for the water opacity.  

\item Generally, the elemental abundances retrieved using ``free chemistry'' models are sub-solar to solar.  (The only exception are the oxygen abundances derived from NIRISS, which are super-solar.)  However, their exact values are model- and mode-dependent.  The elemental abundances inferred using the random forest method trained on the \cite{Crossfield2023ApJ...952L..18C} model grid are super-solar (and more consistent with the findings of the ERS papers).

\item The retrieved C/O values range from sub- to super-solar.  C/O ratios retrieved from the NIRCam spectrum are consistent with the stellar value, because these retrievals are essentially cloud-free.  By contrast, the C/O ratios retrieved from the NIRSpec PRISM spectrum are super- and sub-solar when grey and non-grey clouds are assumed, respectively.

\item The detection of sulphur dioxide from the NIRSpec G395H and PRISM spectra depends on whether grey or non-grey clouds are assumed, because the cloud spectral continuum interacts with the water opacity to influence whether the \ch{SO2} feature is needed to fit the data.

\end{itemize}

\begin{acknowledgements}
A.L. acknowledges partial financial support from the Swiss National Science Foundation and the European Research Council (via a Consolidator Grant to KH; grant number 771620), as well as administrative support from the Centre for Space and Habitability (CSH). A.N. acknowledges financial support from the Coordination of Improvement of Higher Education Personnel (CAPES) and LMU-Munich, and Luan Ghezzi for support and discussion. We thank Daniel Kitzmann and Jo Barstow for their valuable input. This study would not have been possible without the open-source codes \texttt{Helios-r2} \citep{Kitzmann2020ApJ...890..174K} and \texttt{HELA} \citep{Marquez-Neila2018NatAs...2..719M}. Both of them can be found on the Exoclimes Simulation Platform: \href{https://github.com/exoclime}{https://github.com/exoclime}. Additionally, we gratefully acknowledge the open-source libraries in the Python programming language that made this work possible: \textbf{scikit.learn} \citep{scikit-learn2011}, \textbf{numpy} \citep{Harris2020Natur.585..357H}, \textbf{matplotlib} \citep{Hunter2007CSE.....9...90H}, and \textbf{astropy} \citep{astropy:2013, astropy:2018, astropy:2022}.
This publication makes use of The Data \& Analysis Center for Exoplanets (DACE), which is a facility based at the University of Geneva (CH) dedicated to extrasolar planets data visualisation, exchange, and analysis. DACE is a platform of the Swiss National Centre of Competence in Research (NCCR) PlanetS, federating the Swiss expertise in exoplanet research. The DACE platform is available at \href{https://dace.unige.ch}{https://dace.unige.ch}.
\end{acknowledgements}

\bibliographystyle{aa}
\bibliography{references.bib}

\onecolumn
\begin{appendix}

\section{\texttt{Helios-r2} and \texttt{HELA} outcomes}
\label{apx:outcomes}

As explained in the main body of the paper, the ``simplified models'' are nested-sampling retrievals performed with the minimal set of chemical species required to fit the data.  Table \ref{tab:heliosr2_posteriors} states the median values of the posterior distributions of parameters, as well as their 1-$\sigma$ uncertainties.  Elemental abundances (in terms of their solar values) retrieved using random forest retrievals are stated in Table \ref{tab:hela_posteriors}.  Again, these are the median values of the posterior distributions and their 1-$\sigma$ uncertainties.

\begin{table*}[ht]
\caption{Summary of our best-fit \texttt{Helios-r2} retrieval outcomes for each of the four JWST spectra: NIRCam, NIRISS, NIRSpec G395H, and NIRSpec PRISM.}
\label{tab:heliosr2_posteriors}
\centering
\begin{tabular}{ccccccccc}
\hline\hline
\shortstack{ \vspace{0.3cm} \\ Parameter \\ \vspace{0.3cm} } & \shortstack{NIRCam \\ grey \\ 3 FP} & \shortstack{NIRCam \\ non-grey\\ 3 FP} & \shortstack{NIRISS \\ grey\\ 2 FP} & \shortstack{NIRISS \\ non-grey\\ 1 FP} & \shortstack{G395H \\ grey\\ 1 FP} & \shortstack{G395H \\ non-grey\\ 2 FP} & \shortstack{PRISM \\ grey\\ 1 FP} & \shortstack{PRISM \\ non-grey\\ 1 FP} \\
\hline
$\log g$ [cm/s$^2$] & $2.70_{-0.03}^{+0.02}$ & $2.69_{-0.03}^{+0.02}$ & $2.64_{-0.04}^{+0.03}$ & $2.66_{-0.03}^{+0.03}$ & $2.69_{-0.04}^{+0.04}$ & $2.67_{-0.03}^{+0.03}$ & $2.65_{-0.04}^{+0.04}$ & $2.77_{-0.02}^{+0.02}$ \\ \hline
$R_{\rm P}$ [R$_\mathrm{Jup}$] & $1.28_{-0.02}^{+0.02}$ & $1.28_{-0.02}^{+0.02}$ & $1.28_{-0.02}^{+0.02}$ & $1.28_{-0.02}^{+0.02}$ & $1.31_{-0.03}^{+0.03}$ & $1.28_{-0.02}^{+0.02}$ & $1.27_{-0.02}^{+0.02}$ & $1.29_{-0.02}^{+0.02}$ \\ \hline
$R_{\rm S}$ [R$_\odot$] & $0.95_{-0.01}^{+0.01}$ & $0.95_{-0.01}^{+0.01}$ & $0.94_{-0.02}^{+0.02}$ & $0.94_{-0.02}^{+0.02}$ & $0.94_{-0.02}^{+0.02}$ & $0.94_{-0.02}^{+0.01}$ & $0.95_{-0.02}^{+0.01}$ & $0.95_{-0.01}^{+0.01}$ \\ \hline
$\log$ \ch{H2O} & $-4.15_{-0.38}^{+0.49}$ & $-4.10_{-0.41}^{+0.49}$ & $-1.45_{-0.20}^{+0.19}$ & $-1.13_{-0.12}^{+0.08}$ & $-5.28_{-0.36}^{+0.37}$ & $-6.35_{-0.33}^{+0.34}$ & $-3.10_{-0.19}^{+0.20}$ & $-3.53_{-0.31}^{+0.44}$ \\ \hline
$\log$ \ch{CO} & - & - & - & - & - & - & $-2.85_{-0.28}^{+0.17}$ & - \\ \hline
$\log$ \ch{CO2} & $-3.34_{-0.53}^{+0.44}$ & $-3.30_{-0.52}^{+0.43}$ & - & - & $-6.46_{-0.37}^{+0.44}$ & $-6.49_{-0.28}^{+0.34}$ & $-4.34_{-0.22}^{+0.24}$ & $-5.01_{-0.35}^{+0.53}$ \\ \hline
$\log$ \ch{H2S} & - & - & - & - & - & - & $-4.07_{-0.31}^{+0.27}$ & $-4.75_{-0.38}^{+0.42}$ \\ \hline
$\log$ \ch{K} & - & - & $-7.50_{-0.40}^{+0.41}$ & $-5.59_{-0.51}^{+0.46}$ & - & - & $-7.01_{-1.86}^{+1.87}$ & $-7.80_{-1.44}^{+1.43}$ \\ \hline
$\log$ \ch{Na} & - & - & - & - & - & - & $-3.75_{-0.38}^{+0.37}$ & $-5.15_{-0.35}^{+0.44}$ \\ \hline
$\log$ \ch{SO2} & - & - & - & - & - & $-7.55_{-0.28}^{+0.18}$ & $-5.68_{-0.62}^{+0.31}$ & - \\ \hline
$\log$ \ch{CH4} & - & - & - & - & - & - & - & - \\ \hline
$\log \tau_{\mathrm{cloud}}$ & $-2.45_{-1.55}^{+1.75}$ & $-2.08_{-1.82}^{+2.03}$ & $1.34_{-1.06}^{+1.06}$ & $0.66_{-0.15}^{+0.13}$ & $-2.95_{-1.31}^{+2.31}$ & $1.17_{-0.29}^{+0.18}$ & $-0.39_{-0.50}^{+2.23}$ & $2.63_{-0.36}^{+0.24}$ \\ \hline
$\log{P_\mathrm{cloudtop}}$ [bar] & $-1.76_{-2.66}^{+1.90}$ & $-2.19_{-2.16}^{+1.93}$ & $-1.73_{-0.14}^{+0.13}$ & $-2.79_{-0.18}^{+0.22}$ & $-1.44_{-2.96}^{+1.79}$ & $-5.06_{-0.61}^{+0.73}$ & $-1.91_{-0.99}^{+0.38}$ & $-1.62_{-0.24}^{+0.19}$ \\ \hline
Q$_0$ & - & $47.96_{-28.03}^{+31.34}$ & - & $49.31_{-30.66}^{+31.21}$ & - & $49.74_{-30.20}^{+30.84}$ & - & $50.32_{-30.89}^{+30.65}$ \\ \hline
$a_0$ & - & $4.43_{-0.86}^{+0.95}$ & - & $5.61_{-0.48}^{+0.28}$ & - & $5.49_{-0.62}^{+0.35}$ & - & $3.57_{-0.37}^{+0.54}$ \\ \hline
$\log{r_{\mathrm{cloud}}}$ [cm] & - & $-3.86_{-1.85}^{+1.74}$ & - & $-3.99_{-1.89}^{+1.81}$ & - & $-4.06_{-1.91}^{+1.90}$ & - & $-3.92_{-1.81}^{+1.78}$ \\ \hline
\end{tabular}
\tablefoot{Each simplified model includes uniform abundance profiles for the represented species. Best-fits are presented for both grey and non-grey cloud models. The number of free parameters for each temperature-pressure profile is denoted as FP.}
\end{table*}

\begin{table*}[ht]
\caption{Inferred elemental abundances (in terms of their solar values) from our best-fit \texttt{HELA} retrieval outcomes for each of the four JWST spectra: NIRCam, NIRISS, NIRSpec G395H, and NIRSpec PRISM.}
\label{tab:hela_posteriors}
\centering
\begin{tabular}{ccccc}
\hline\hline
Parameter & NIRCam & NIRISS & NIRSpec G395H & NIRSpec PRISM \\
\hline
$\log$ C &
$0.748_{-0.493}^{+0.366}$ &
$1.110_{-0.637}^{+0.634}$ &
$0.748_{-0.493}^{+0.507}$ &
$1.000_{-0.745}^{+0.748}$ \\ \hline
$\log$ O &
$0.875_{-0.127}^{+0.239}$ &
$2.000_{-0.252}^{+0.000}$ &
$1.260_{-0.255}^{+0.493}$ &
$1.480_{-0.477}^{+0.523}$ \\ \hline
$\log$ S &
$1.000_{-0.745}^{+0.725}$ &
$1.000_{-0.745}^{+0.748}$ &
$0.748_{-0.493}^{+0.366}$ &
$0.748_{-0.493}^{+0.366}$ \\ \hline
\end{tabular}
\end{table*}

\section{\texttt{Helios-r2} posteriors and spectra}
\label{apx:PosteriorsSpectra}

\begin{figure*}[ht]
    \centering
    \includegraphics[width=0.95\textwidth]{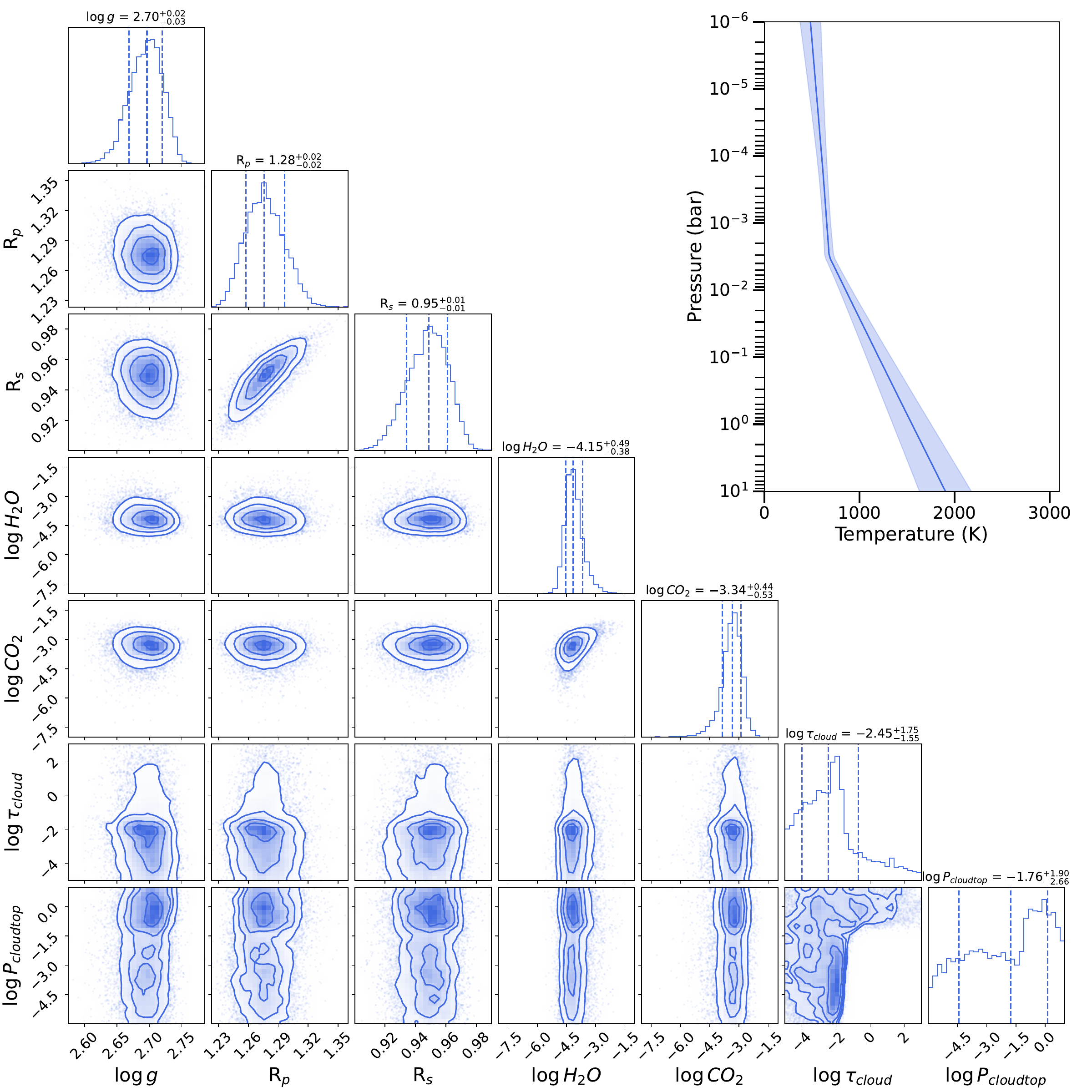}
    \vfill
    \includegraphics[width=\textwidth]{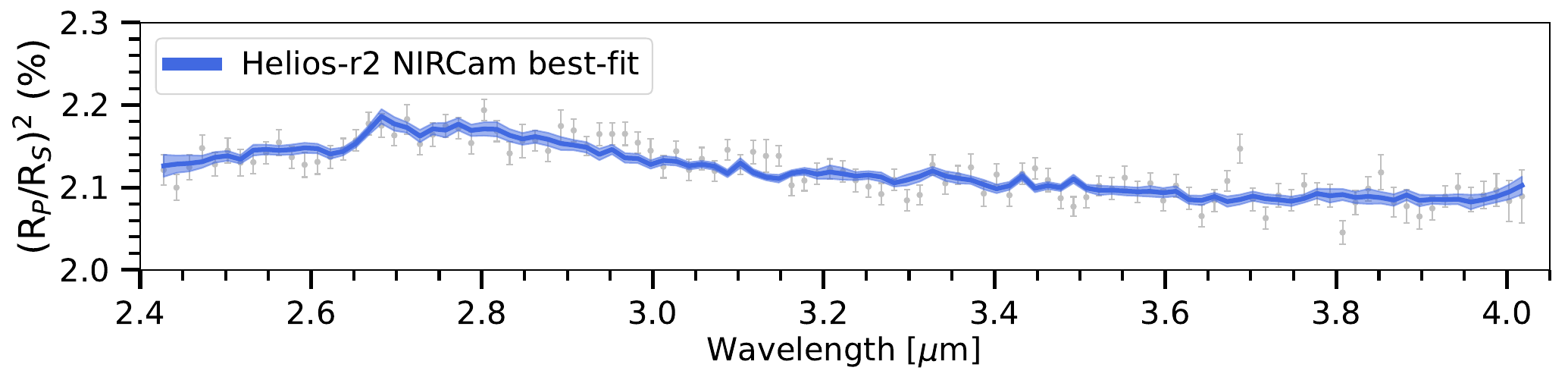}
    \caption{Full set of posterior distributions from our nested sampling retrieval performed on the NIRCam spectrum.  A simplified model assuming grey clouds was used.  Relative to the model assuming non-grey clouds (see Fig.~\ref{fig:NIRCam_Corner_unfav}), the logarithm of the Bayes factor is 0.41.  For each marginalised posterior distribution, the vertical dashed lines indicate the median value of the distribution and the associated 1-$\sigma$ uncertainties.  The temperature-pressure profile also displays median values and the associated 1-$\sigma$ uncertainties.  The best-fit spectrum is associated with 1-$\sigma$ uncertainties as well.}
    \label{fig:NIRCam_Corner_fav}
\end{figure*}

\begin{figure*}[ht]
    \centering
    \includegraphics[width=\textwidth]{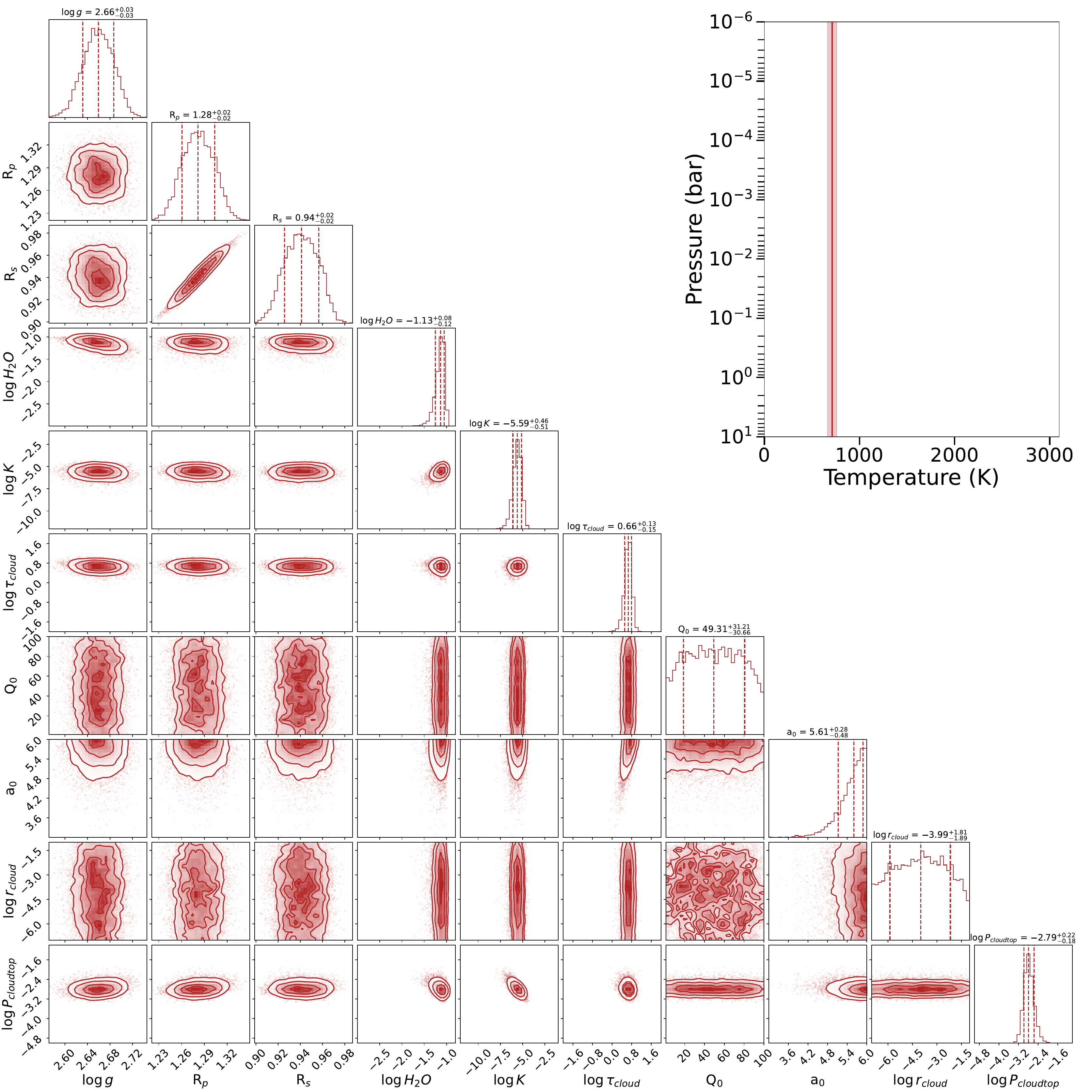}
    \vfill
    \includegraphics[width=\textwidth]{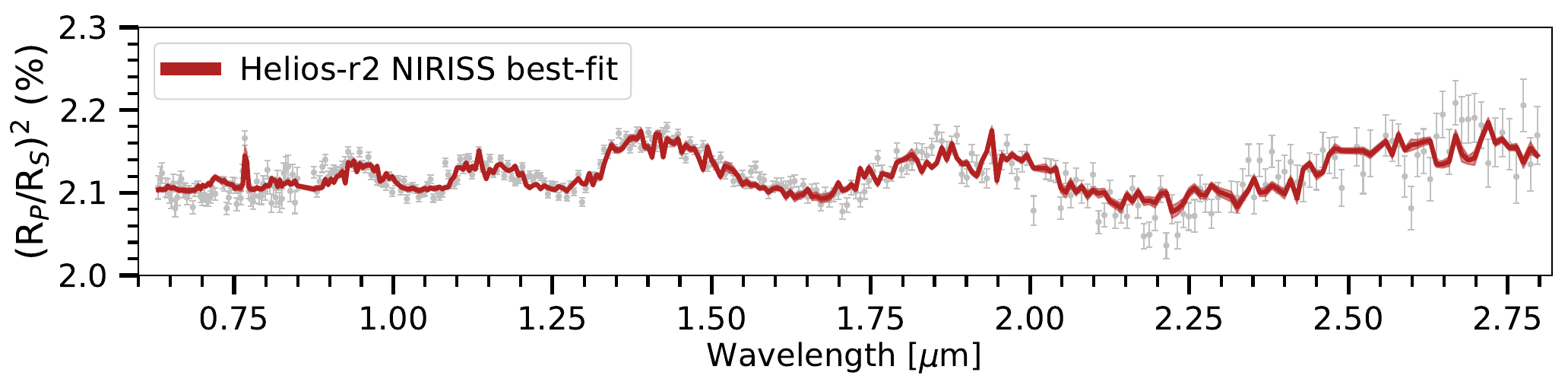}
    \caption{Same as Fig. \ref{fig:NIRCam_Corner_fav} but for the NIRISS spectrum and assuming the non-grey cloud model.  Relative to the model assuming grey clouds (see Fig.~\ref{fig:NIRISS_Corner_unfav}), the logarithm of the Bayes factor is 11.}
    \label{fig:NIRISS_Corner_fav}
\end{figure*}

\begin{figure*}[ht]
    \centering
    \includegraphics[width=\textwidth]{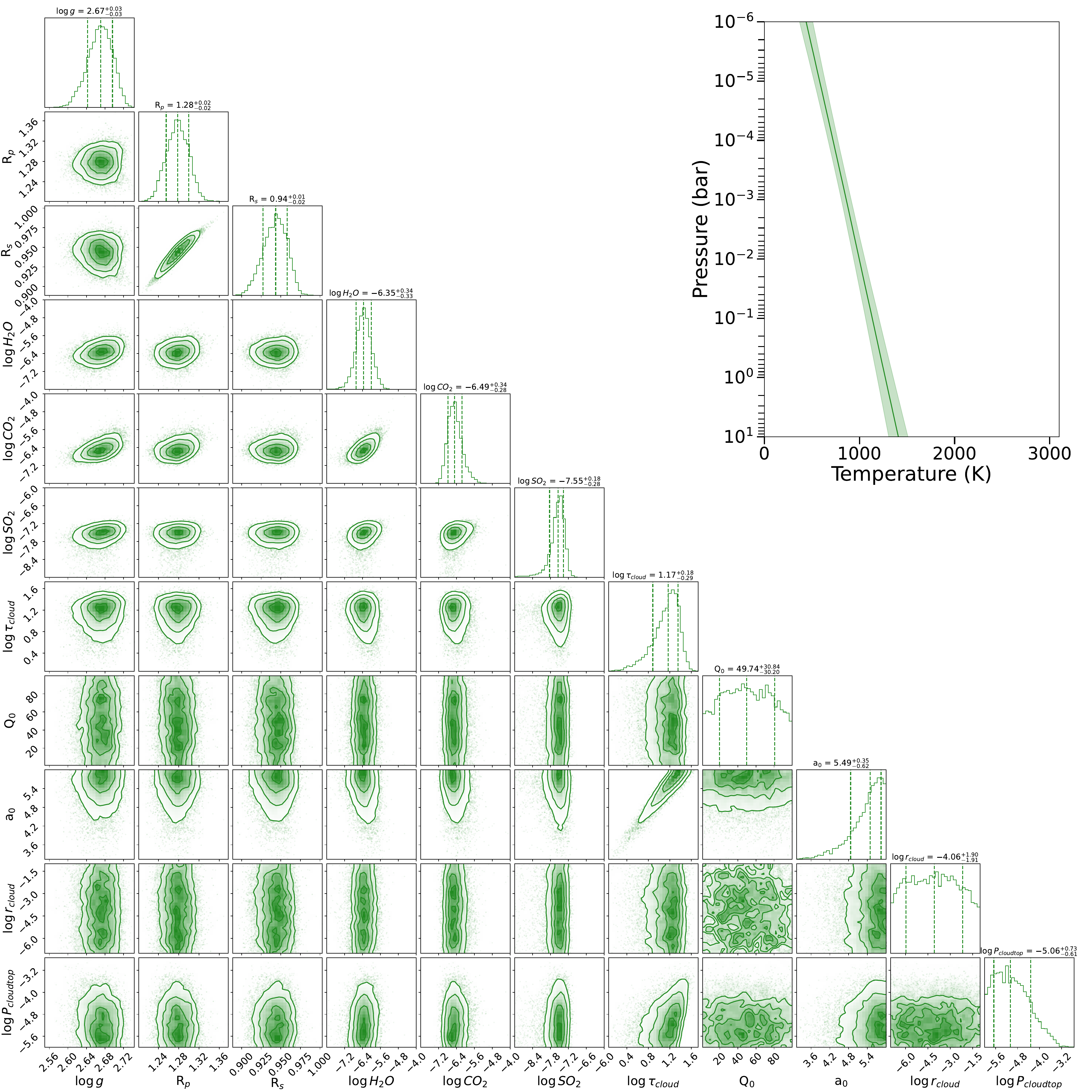}
    \vfill
    \includegraphics[width=\textwidth]{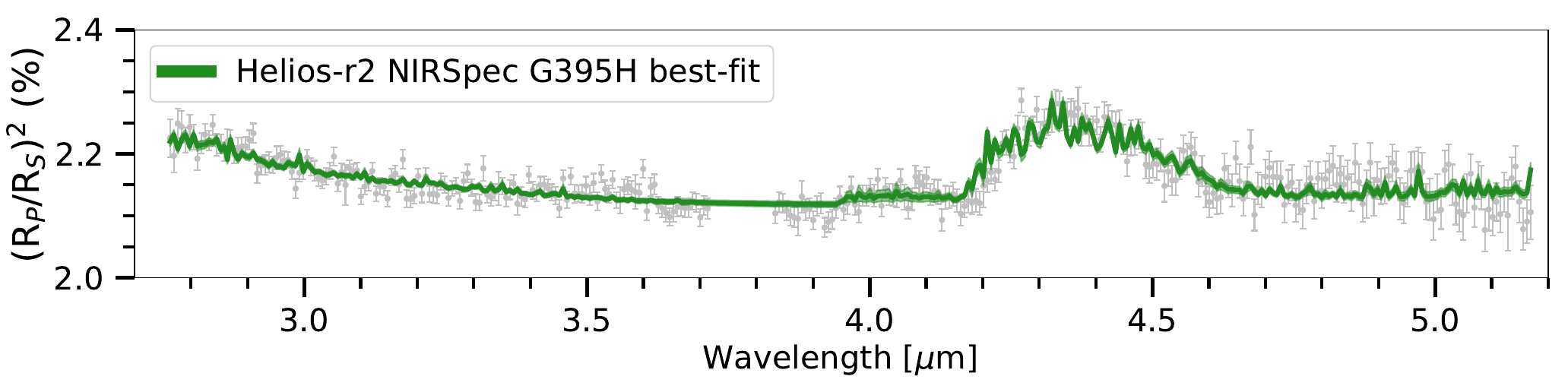}
    \caption{Same as Fig. \ref{fig:NIRCam_Corner_fav} but for the NIRSpec G395H spectrum and assuming the non-grey cloud model.  Relative to the model assuming grey clouds (see Fig.~\ref{fig:G395H_Corner_unfav}), the logarithm of the Bayes factor is 34.}
    \label{fig:G395H_Corner_fav}
\end{figure*}

\begin{figure*}[ht]
    \centering
    \includegraphics[width=\textwidth]{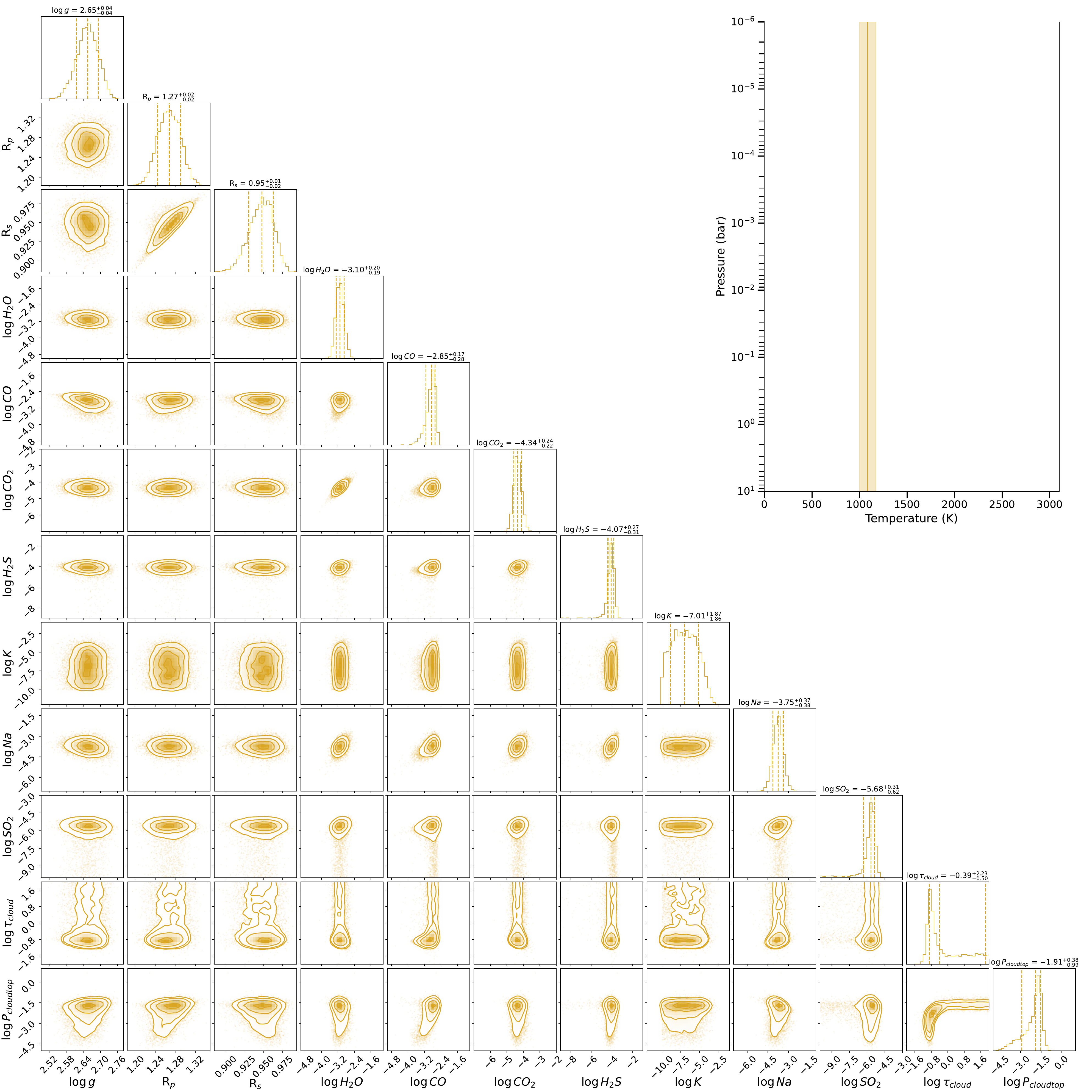}
    \vfill
    \includegraphics[width=\textwidth]{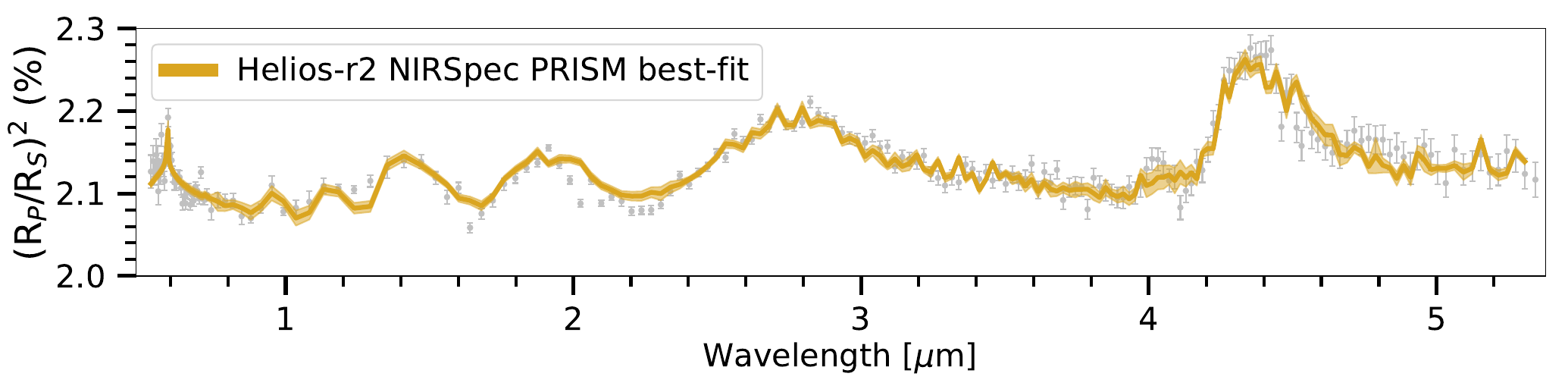}
    \caption{Same as Fig. \ref{fig:NIRCam_Corner_fav} but for the NIRSpec PRISM spectrum and assuming the grey cloud model.  Relative to the model assuming non-grey clouds (see Fig.~\ref{fig:PRISM_Corner_unfav}), the logarithm of the Bayes factor is 0.61.}
    \label{fig:PRISM_Corner_fav}
\end{figure*}

\begin{figure*}[ht]
    \centering
    \includegraphics[width=\textwidth]{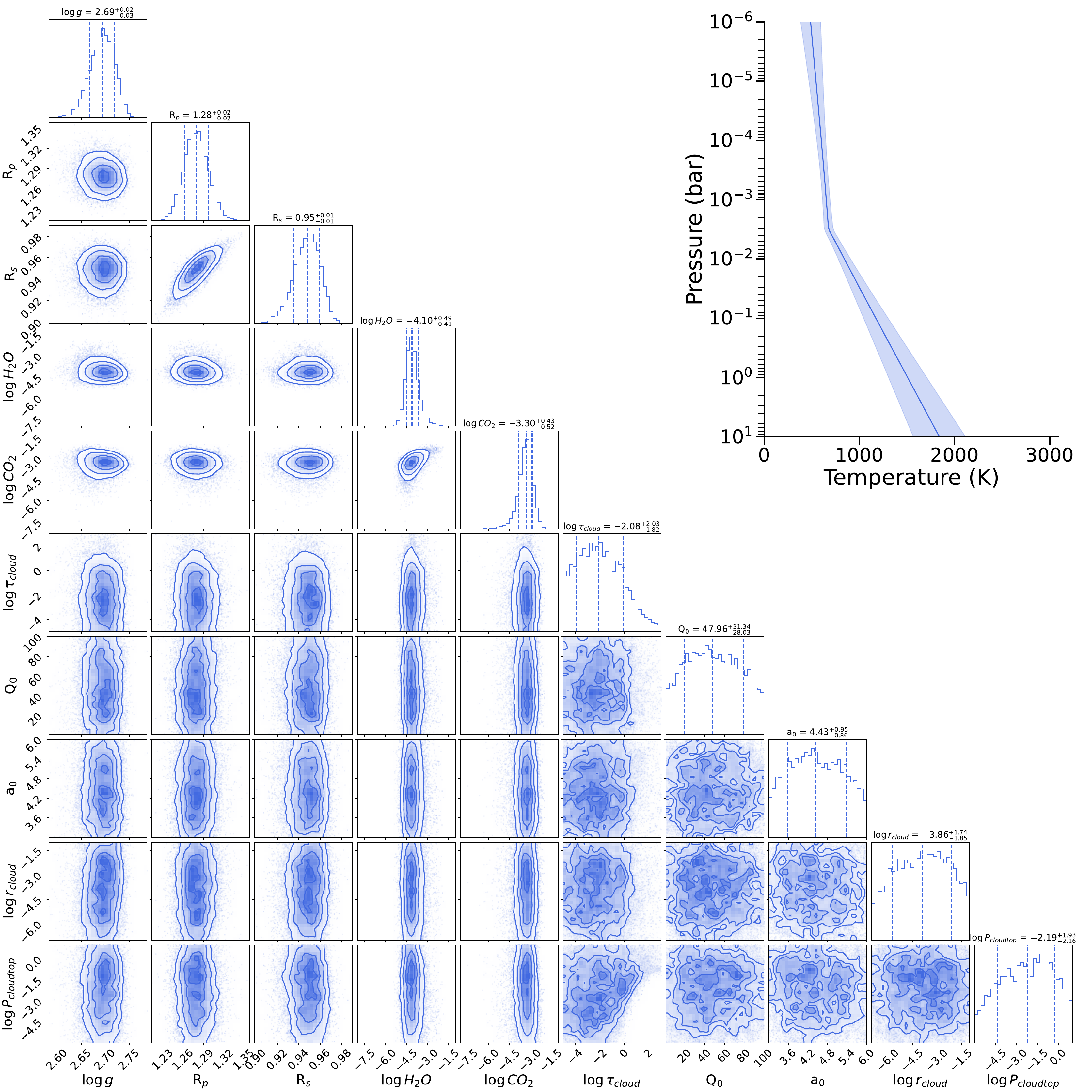}
    \vfill
    \includegraphics[width=\textwidth]{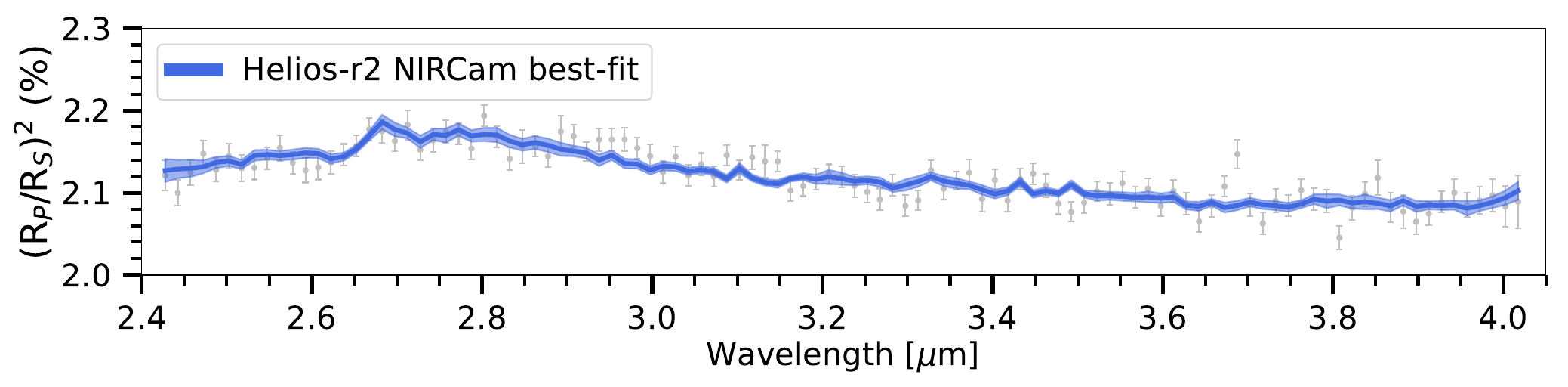}
    \caption{Counterpart to Fig. \ref{fig:NIRCam_Corner_fav} assuming non-grey clouds.}
    \label{fig:NIRCam_Corner_unfav}
\end{figure*}

\begin{figure*}[ht]
    \centering
    \includegraphics[width=\textwidth]{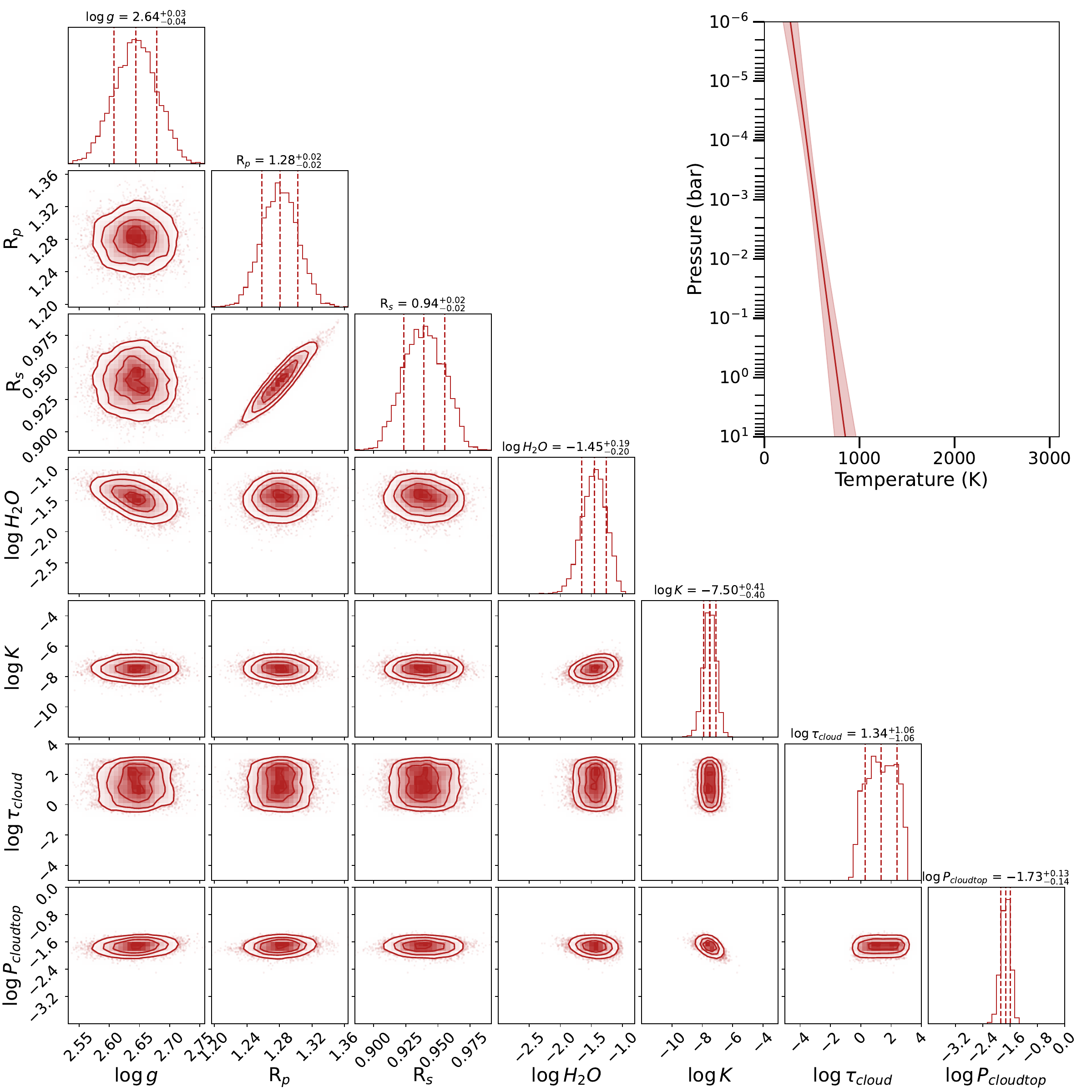}
    \vfill
    \includegraphics[width=\textwidth]{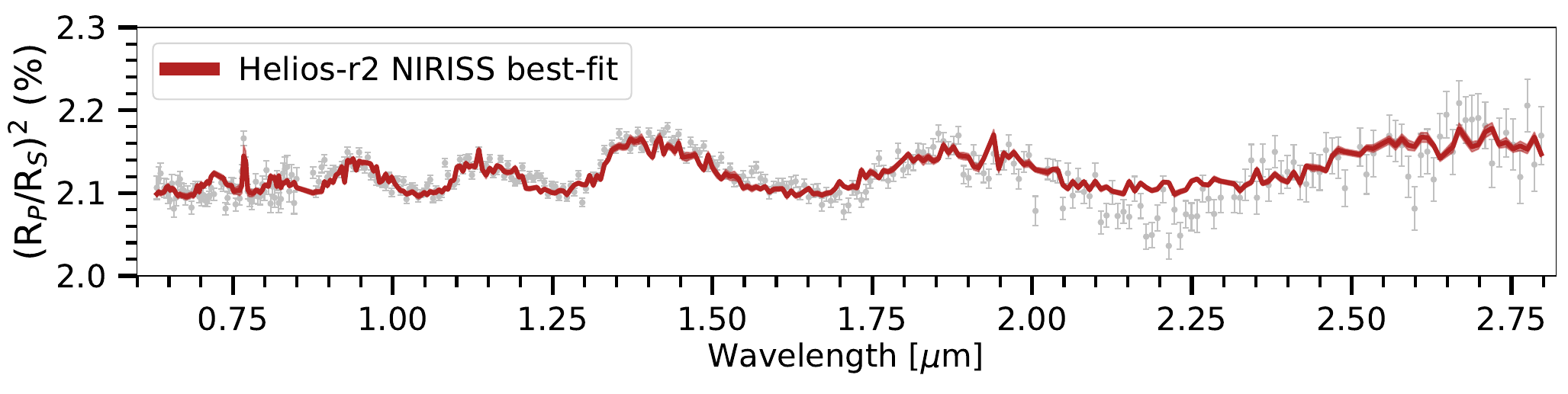}
    \caption{Counterpart to Fig. \ref{fig:NIRISS_Corner_fav} assuming grey clouds.}
    \label{fig:NIRISS_Corner_unfav}
\end{figure*}

\begin{figure*}[ht]
    \centering
    \includegraphics[width=\textwidth]{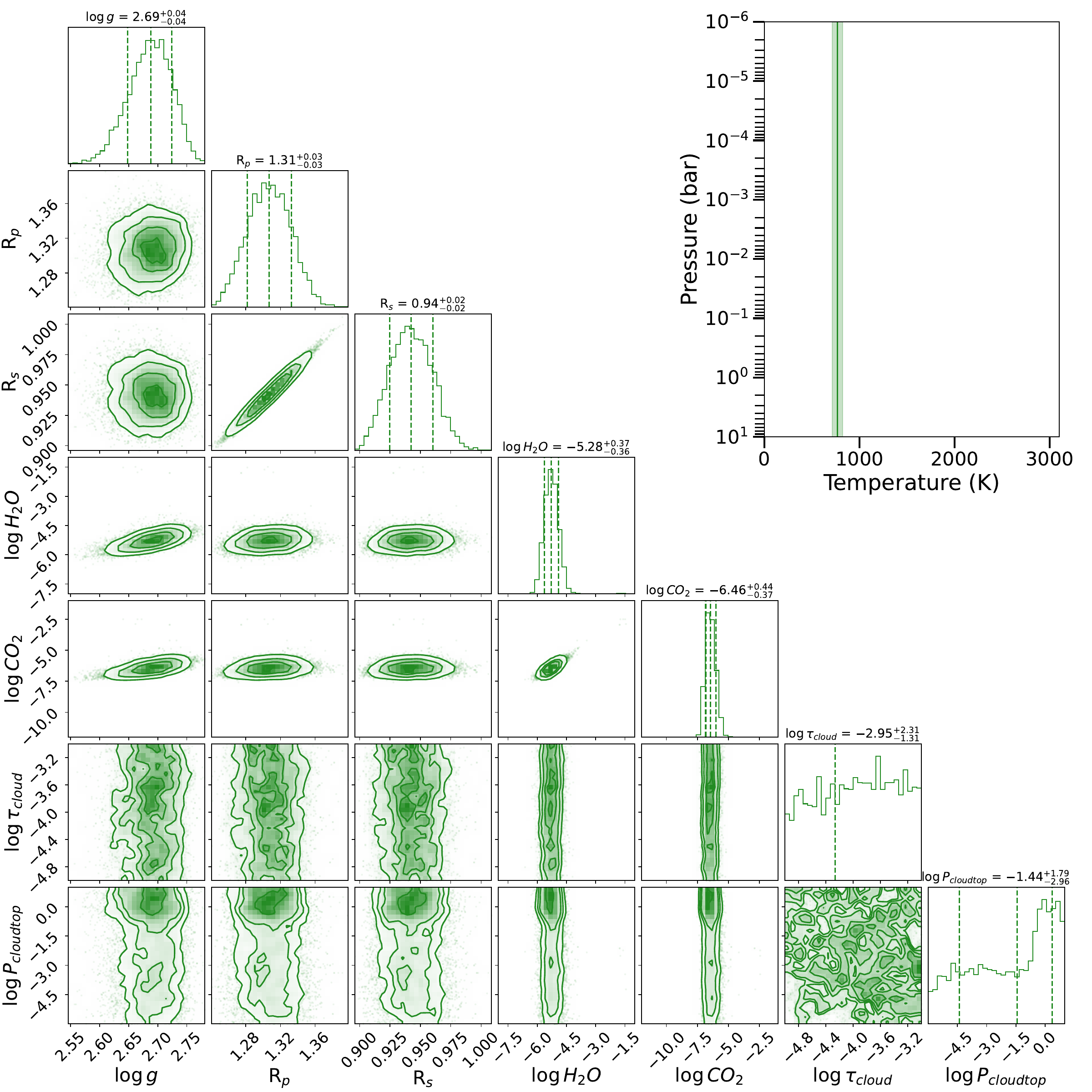}
    \vfill
    \includegraphics[width=\textwidth]{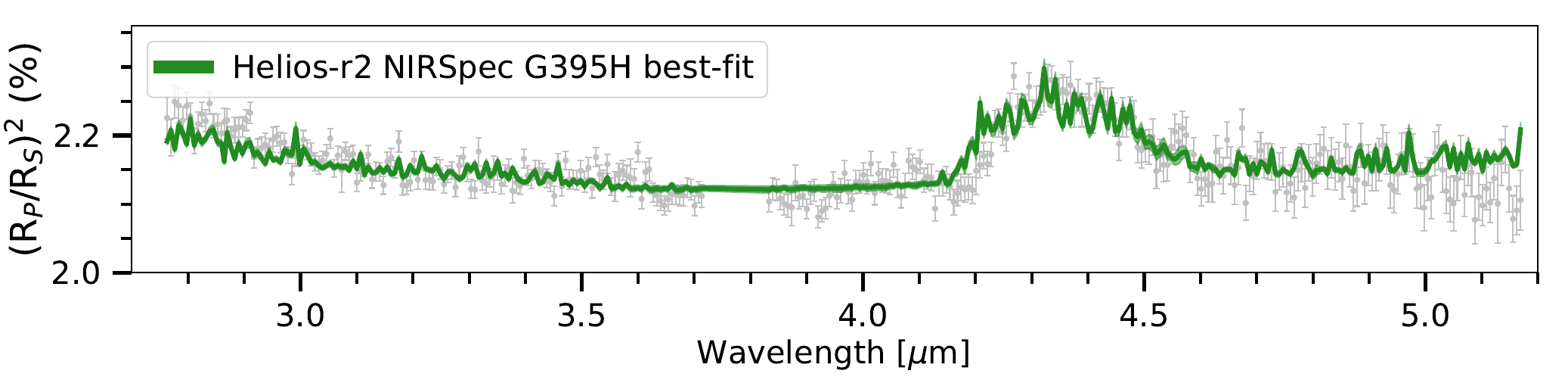}
    \caption{Counterpart to Fig. \ref{fig:G395H_Corner_fav} assuming grey clouds.}
    \label{fig:G395H_Corner_unfav}
\end{figure*}

\begin{figure*}[ht]
    \centering
    \includegraphics[width=\textwidth]{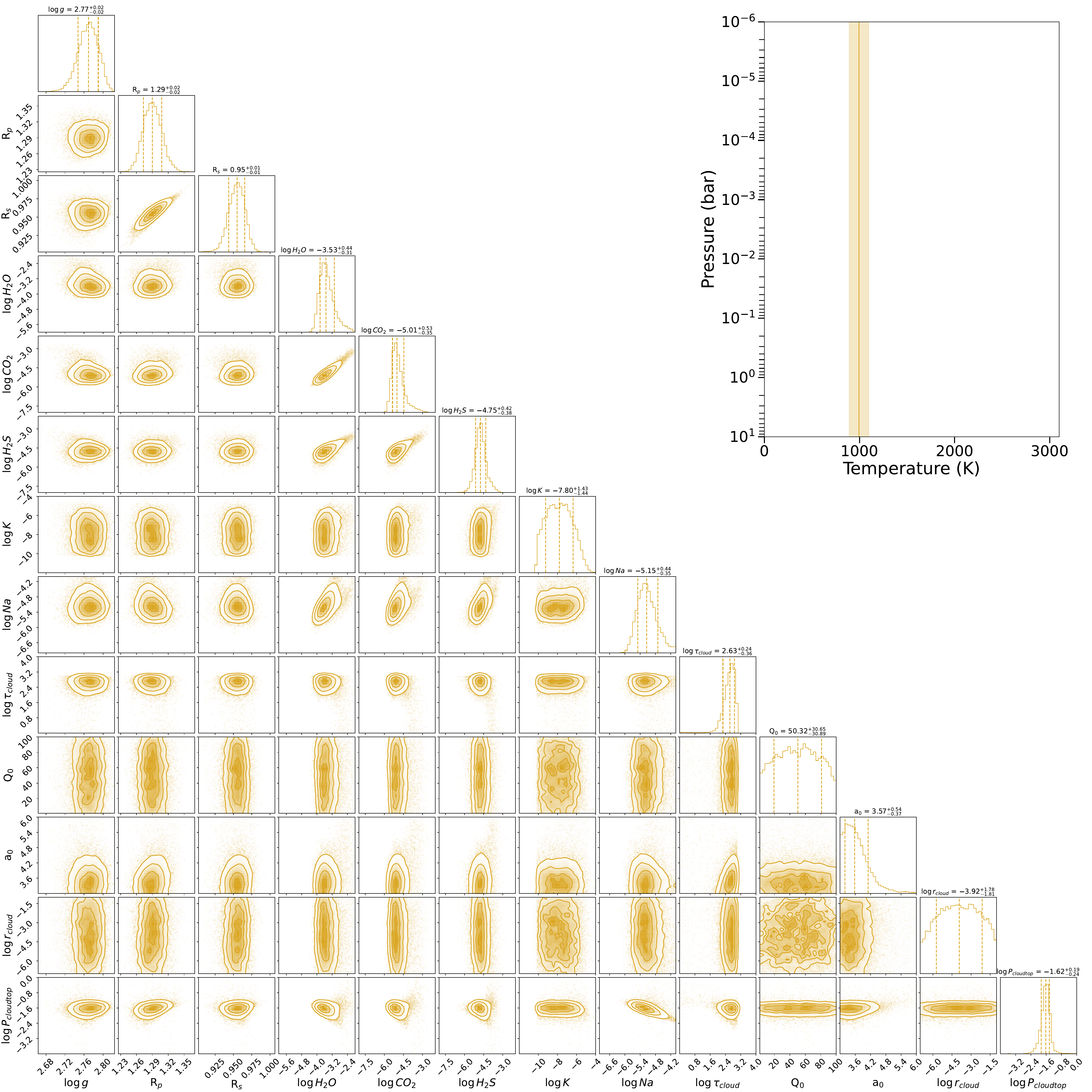}
    \vfill
    \includegraphics[width=\textwidth]{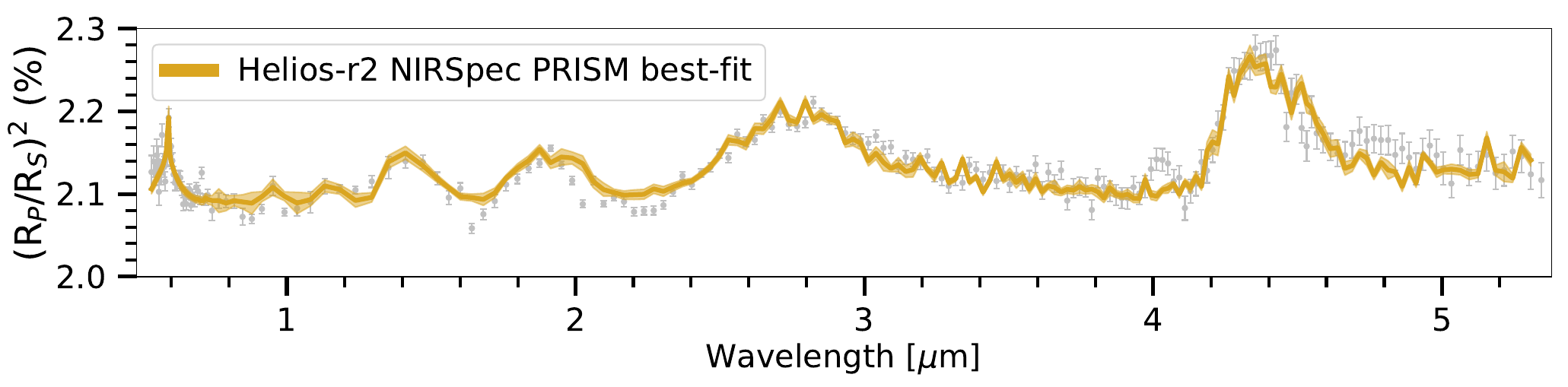}
    \caption{Counterpart to Fig. \ref{fig:PRISM_Corner_fav} assuming non-grey clouds.}
    \label{fig:PRISM_Corner_unfav}
\end{figure*}

For completeness, the full sets of posterior distributions of the parameters and spectra from the simplified \texttt{Helios-r2} (nested sampling) retrieval models are shown in Figs.~\ref{fig:NIRCam_Corner_fav} to \ref{fig:PRISM_Corner_unfav}.

\FloatBarrier
\section{\texttt{HELA} posteriors}
\label{apx:PosteriorsHELA}

\begin{figure*}[ht!]
    \centering
    \includegraphics[width=0.55\textwidth]{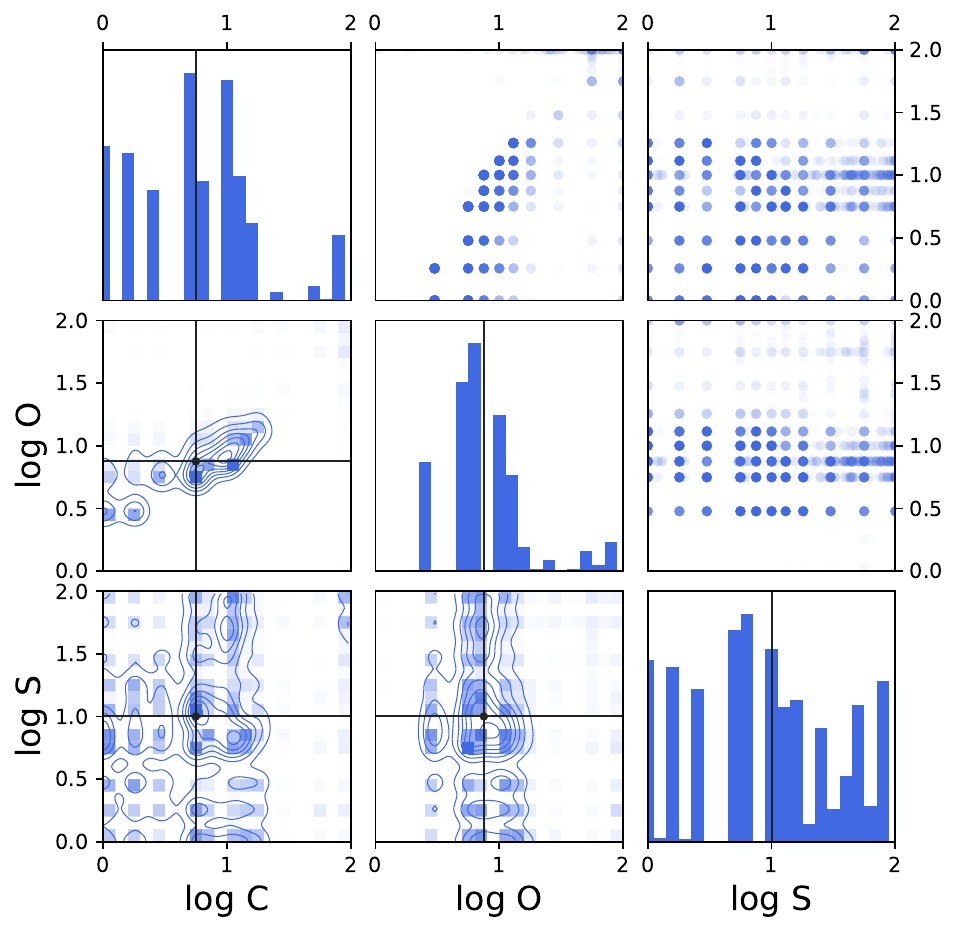}
    \caption{Full set of posterior distributions from our random forest retrieval (trained on the model grid of \citealt{Crossfield2023ApJ...952L..18C}) performed on the NIRCam spectrum. Shown are the logarithm (base 10) of the elemental abundances of C, O, and S, in terms of their solar value. In each scatter plot, every point represents an individual prediction generated by a single regression tree within the random forest ensemble. The solid vertical lines denote the median values of the posterior distributions.}
    \label{fig:HELA_posteriors_NIRCam}
\end{figure*}

\begin{figure*}[ht!]
    \centering
    \includegraphics[width=0.55\textwidth]{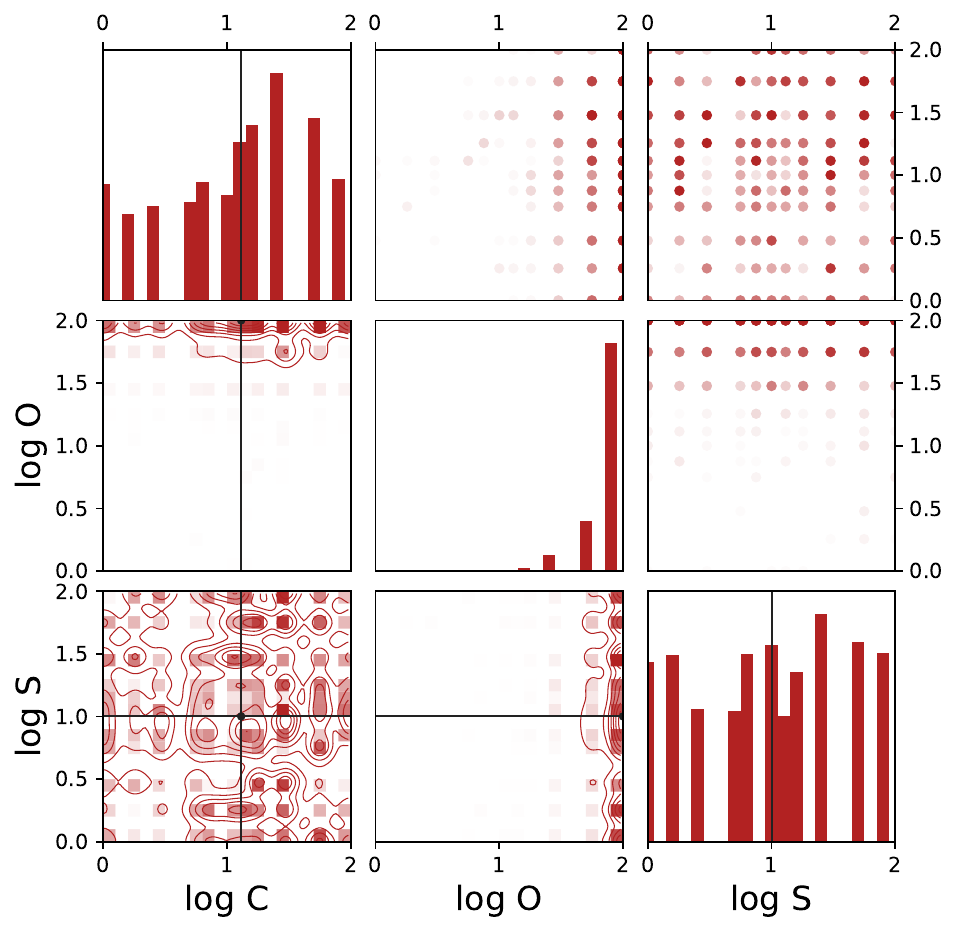}
    \caption{Same as Fig. \ref{fig:HELA_posteriors_NIRCam} but for the NIRISS spectrum.}
    \label{fig:HELA_posteriors_NIRISS}
\end{figure*}

\begin{figure*}[ht!]
    \centering
    \includegraphics[width=0.55\textwidth]{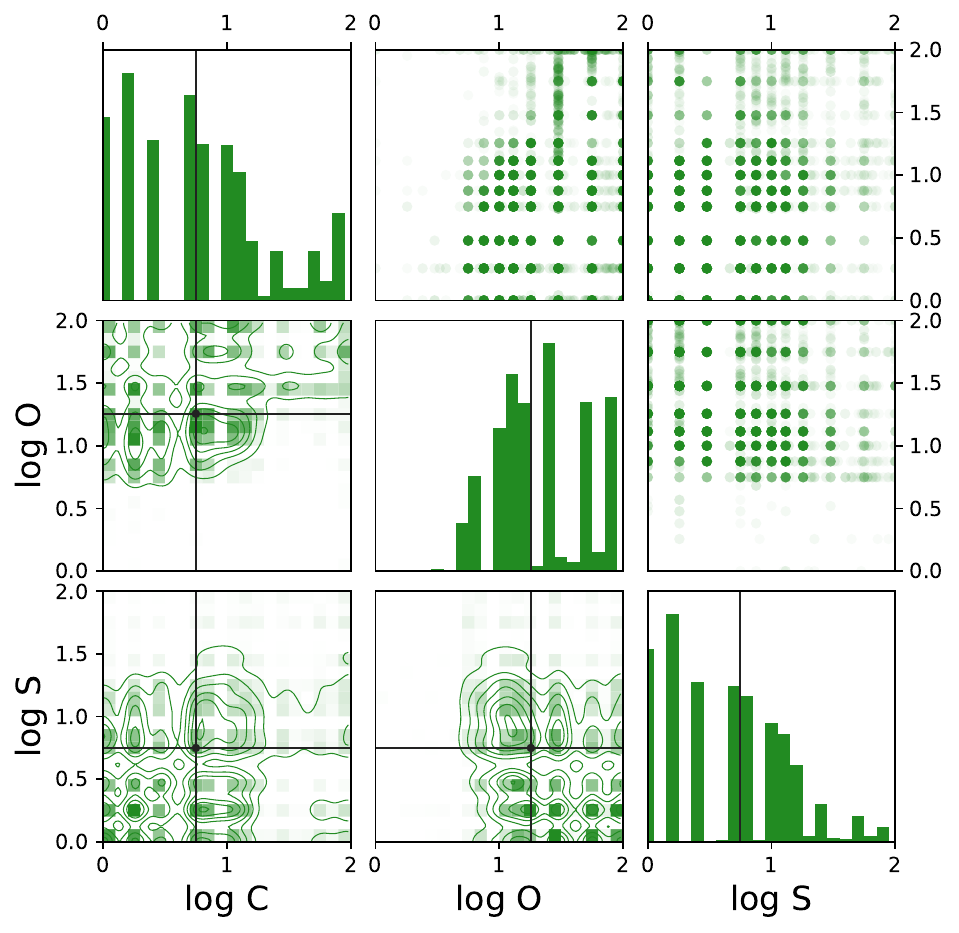}
    \caption{Same as Fig. \ref{fig:HELA_posteriors_NIRCam} but for the NIRSpec G395H spectrum.}
    \label{fig:HELA_posteriors_G395H}
\end{figure*}

\begin{figure*}[ht]
    \centering
    \includegraphics[width=0.55\textwidth]{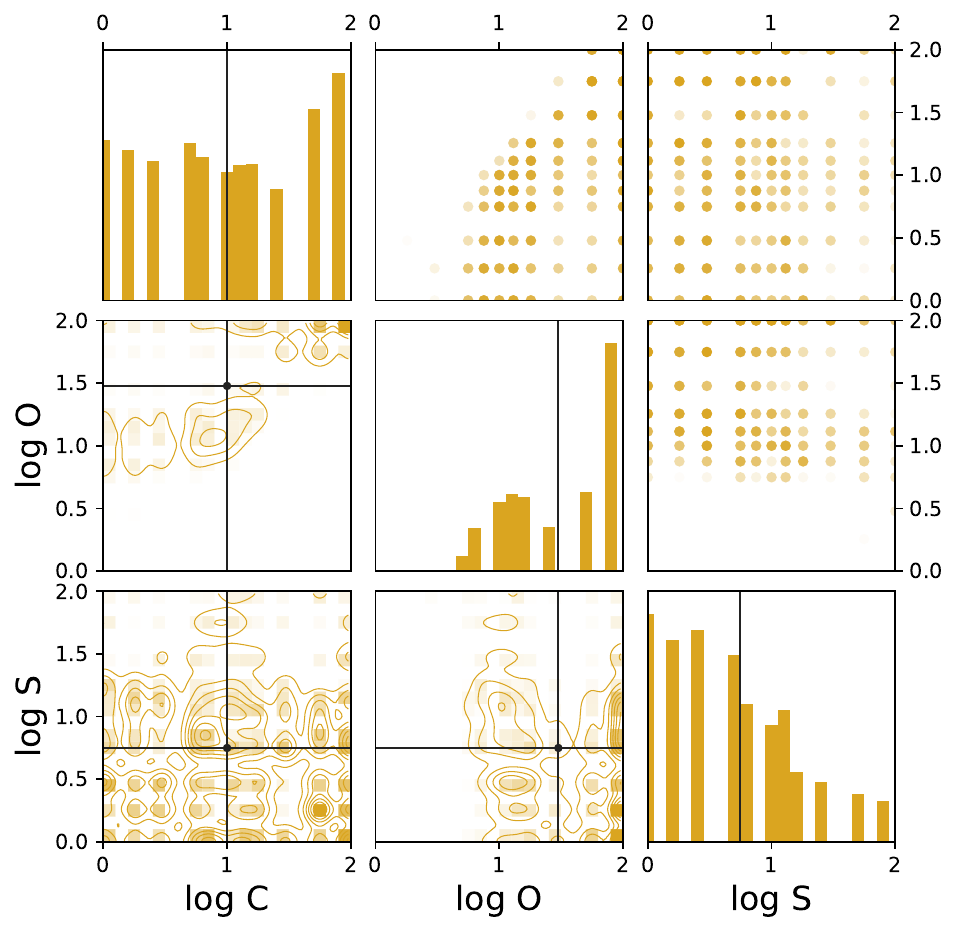}
    \caption{Same as Fig. \ref{fig:HELA_posteriors_NIRCam} but for the NIRSpec PRISM spectrum.}
    \label{fig:HELA_posteriors_PRISM}
\end{figure*}

For completeness, the full sets of posterior distributions of the parameters from the \texttt{HELA} (random forest) retrieval models are shown in Figs.~\ref{fig:HELA_posteriors_NIRCam} to \ref{fig:HELA_posteriors_PRISM}. The random forest algorithm was trained on the model grid of \cite{Crossfield2023ApJ...952L..18C}.

\end{appendix}

\end{document}